\documentclass[12pt,journal,draftclsnofoot,onecolumn,english]{IEEEtran}
\usepackage{comment}
\usepackage{epsfig}
\usepackage{times}
\usepackage{float}
\usepackage{afterpage}
\usepackage{multirow}
\usepackage{algorithm}
\usepackage{makecell}
\usepackage{algpseudocode}
\usepackage{amsmath}
\usepackage{amstext,cite}
\usepackage{amssymb,bm}
\usepackage{latexsym}
\usepackage{color}
\usepackage{graphicx}
\usepackage{amsmath}
\usepackage{amsthm}
\usepackage{graphicx}
\usepackage[center]{caption}
\usepackage{pstricks}
\usepackage{caption}
\usepackage{subcaption}
\usepackage{booktabs}
\usepackage{multicol}
\usepackage{lipsum}
\usepackage[T1]{fontenc}
\usepackage{hyperref}
\usepackage{aecompl}
\usepackage{bbm}
\usepackage{empheq}
\usepackage{mathrsfs}
\allowdisplaybreaks
\usepackage{etoolbox}
\usepackage{setspace}
\usepackage{cite}
\usepackage{xcolor}
\usepackage{epsfig}
\usepackage{textcomp}
\usepackage{indentfirst}
\usepackage{xcolor}
\usepackage{times}
\usepackage{float}
\AtBeginEnvironment{algorithmic}{\setstretch{1.5}}
\usepackage{algorithm}
\theoremstyle{definition}
\newtheorem{definition}{Definition}
\newtheorem{example}{Example}

\newtheorem{remark}{Remark}
\newtheorem{lemma}{Lemma}

\begin{document}
\title{A Low-Complexity Framework for Multi-access Coded Caching  with Arbitrary User-cache Access Topology}

\author{
Ting~Yang, 
Kai~Wan,~\IEEEmembership{Member,~IEEE,} 
Minquan~Cheng,~\IEEEmembership{Member,~IEEE,}  
Xinping~Yi,~\IEEEmembership{Member,~IEEE,}
Robert~Caiming~Qiu,~\IEEEmembership{Fellow,~IEEE,}
and~Giuseppe~Caire,~\IEEEmembership{Fellow,~IEEE}
\thanks{
T.~Yang,  K.~Wan, and R.~C.~Qiu are with the School of Electronic Information and Communications,
Huazhong University of Science and Technology, 430074  Wuhan, China,  (e-mail: \{yangting, kai\_wan,caiming\}@hust.edu.cn).
}
\thanks{M.~Cheng is with the Key Laboratory of Education Blockchain and
Intelligent Technology, Ministry of Education, and Guangxi Key Laboratory of
Multi-Source Information Mining and Security, Guangxi Normal University,
541004 Guilin, China (e-mail: chengqinshi@hotmail.com).}
\thanks{
X.~Yi is with the Southeast University, Nanjing 210096, China, (e-mail:xyi@seu.edu.cn).}
\thanks{G.~Caire is with the Electrical Engineering and Computer Science Department, Technische Universit\"at Berlin, 10587 Berlin, Germany (e-mail: caire@tu-berlin.de). 
}
}

\maketitle
\begin{abstract}
This paper studies the multi-access coded caching (MACC) problem with arbitrary user-cache access topology, which extends existing MACC models that rely on highly structured and combinatorially designed topologies. We consider a MACC system consisting of a single server, $\Lambda$ cache-nodes, and $K$ user-nodes. The server stores $N$ equal-size files, each cache-node has a storage capacity of $M$ files, and each user-node $k\in[K]$ can access an arbitrary subset of cache-nodes $\mathcal{A}_k\subseteq[\Lambda]$ and retrieve the cached content stored in cache-nodes $\mathcal{A}_k$. The objective is to design a universal framework for the MACC delivery problem. Decoding conflicts among the requested packets are captured by a conflict graph, and the design of the delivery is reduced to a graph coloring problem, where achieving a lower transmission load corresponds to coloring the graph using fewer colors. Under this formulation, the classical DSatur algorithm achieves a transmission load close to the index-coding (IC) converse bound, thereby providing a practical benchmark. However, its computational complexity becomes prohibitive for large-scale graphs. To overcome this limitation, we develop a learning-driven approach using graph neural networks (GNNs) that efficiently constructs coded multicast transmissions  with performance close to the theoretical bounds and generalizes across different user-cache access topologies and numbers of users. In addition, we extend the IC converse bound to MACC systems with arbitrary access topology and propose a low-complexity greedy approximation that closely matches the IC converse bound. Numerical results demonstrate that the proposed approach achieves performance close to the DSatur algorithm and the IC converse bound, while significantly reducing computational complexity, making it well-suited for large-scale MACC systems.
\end{abstract}


\section{Introduction}
Modern wireless networks suffer from highly time-varying user traffic, which causes congestion during peak hours and inefficient utilization of network resources during off-peak periods.
Caching has been widely recognized as an effective approach to mitigate this imbalance by proactively storing popular content during off-peak times, thereby reducing redundant transmissions and alleviating peak-time traffic loads \cite{paschos2018role}. Coded caching was first introduced by Maddah-Ali and Niesen (MN) in \cite{maddah2014fundamental}. In the MN coded caching model, a single server with a library of $N$ files is connected to $K$ users over a shared-link, where each user is equipped with a local cache of size $M$ files. The system operates in two phases. In the \emph{placement phase}, each user caches some packets of the files without knowledge of future demands. In the \emph{delivery phase}, each user requests one file from the library, and the server broadcasts coded multicast transmissions to satisfy all users’ demands. By jointly designing the placement and delivery phases, the MN scheme enables each multicast transmission to simultaneously serve multiple users, thereby creating a global caching gain. Specifically, when the cache size satisfies $M = \frac{tN}{K}$ for some $t \in \{0,1,\dots,K\}$, the MN scheme achieves a worst-case normalized transmission load given by $\frac{K\left(1-\frac{M}{N}\right)}{1+\frac{KM}{N}}$. Here, the term $1-\frac{M}{N}$ corresponds to the \emph{local caching gain}, representing the fraction of each file not stored in the cache-nodes of users, while the term $1+\frac{KM}{N}$ captures the \emph{coded caching gain}, which quantifies the average number of users simultaneously served by each multicast transmission. For general cache sizes, the achievable memory-load tradeoff can be obtained via memory sharing. Moreover, 
the MN scheme was then shown to be exactly optimal 
under the constraint of uncoded cache placement \cite{wan2020index,exactrateuncoded}.

Converse bounds under coded cache placement were also proposed in the literature. A cut-set converse bound was proposed in \cite{maddah2014fundamental}, with a multiplicative gap to the MN scheme of up to $12$. An improved converse bound was proposed in~\cite{yufactor2TIT2018}, with a multiplicative gap to the MN scheme of up to $2$. In order to obtain a tighter converse bound,   the authors in \cite{tian2018symmetrybound} proposed an approach of computer-aided converse bound that 
computes the converse bound as an entropy linear program (LP) \cite{YeungLP}, which  involves a separate variable for each distinct joint entropy of cache contents and transmitted messages, as well as linear inequalities representing Shannon-type constraints. To reduce the computational complexity of the LP, the   inherent symmetry in the caching problem was considered in \cite{tian2018symmetrybound} to reduce the size of the entropy linear program. 
 However, even after symmetry reduction, the number of entropy variables and constraints still grows rapidly with the number of users $K$, making the approach computationally challenging for large-scale systems.

Considering the design on coded caching schemes, 
the authors in \cite{yan2017placement} introduced a combinatorial framework known as the \emph{placement delivery array} (PDA), which provides a universal characterization of coded caching schemes with uncoded cache placement and clique-cover-based delivery.  
The PDA framework can be naturally extended to accommodate a wide range of general network topologies, including combinatorial networks \cite{cheng2020combination}, device-to-device (D2D) networks \cite{ji2015D2D}, hierarchical caching networks \cite{kong2023hiearchical}, multi-antenna networks \cite{yang2023multiple,lampiris2018adding,naderializadeh2017fundamental,namboodiri2023extended,salehi2021low,niu2024reflecting}, and multi-access caching systems \cite{cheng2021maccpda,sasi2021maccpda,wang2023maccpda,zhang2022-2maccpda}.

\subsection{Multi-access coded caching}

The classical coded caching (shared-link) framework assumes that each user is equipped with a dedicated cache that cannot be accessed by other users. In practical wireless and edge networks, however, caching resources are typically deployed at the network edge, such as base stations, access points, or edge servers. These edge cache-nodes often have larger storage capacities and can be accessed by multiple nearby users at high data rates, which naturally motivates the study of the multi-access coded caching (MACC) problem. The MACC model was first introduced in \cite{hachem2017multi-level}, where cache-nodes store content and cache-less users retrieve their requested data by accessing a subset of cache-nodes according to a given user-cache access topology. Compared with the classical model where each user is equipped with a dedicated cache-node, MACC provides greater flexibility in cache deployment and enables higher spatial reuse of cached content.

Most existing works on MACC focus on highly structured and regular access topologies, typically derived from specific combinatorial constructions. Such structures often impose rigid constraints on the numbers of users and cache-nodes, thereby limiting their applicability to more general and irregular network scenarios. A representative example is the cyclic wrap-around topology studied in \cite{hachem2017multi-level,sasi2021maccpda,reddy2021multiaccess,cheng2021maccpda}, where the numbers of users and cache-nodes are identical and each user accesses a fixed number of neighboring cache-nodes in a cyclic fashion. However, under practical cache size constraints, the achievable coded caching gain of such a cyclic MACC topology is fundamentally limited. To overcome these limitations, several works have proposed MACC schemes based on more general combinatorial structures. In particular, symmetric combinatorial MACC systems with $\Lambda$ cache-nodes and $\binom{\Lambda}{a}$ users, where each user accesses exactly $a$ cache-nodes, were investigated in \cite{brunero2022fundamentalmacc,namboodiri2024macc}. Furthermore, across-resolvable combinatorial designs were considered in \cite{katyal2021macc,das2022macc,muralidhar2021macc}, which construct MACC access topology with $\Lambda = m q$ cache-nodes and $K = \binom{m}{a} q^{a}$ users, where $q = \frac{N}{M} \geq 2$ and $m \geq a$ are positive integers. More recently, a universal coded caching framework based on $t$-designs and $t$-group divisible designs ($t$-GDDs) was proposed in \cite{cheng2025t_designmacc}.  This framework subsumes across-resolvable designs as special cases and enables the construction of coded caching schemes over a broader class of structured MACC topologies.
 However, for arbitrary user-cache access topology, the problem becomes generally difficult due to the lack of regularity and symmetry in the underlying connectivity structure. In particular, once the regularity and symmetry required by existing combinatorial constructions disappear, both the construction of achievability schemes and the derivation of information-theoretic converse bounds must be addressed in an algorithmic way. Existing algorithmic approaches often incur prohibitively high computational complexity, especially for large-scale systems and irregular access topologies. This motivates the development of a low-complexity and general framework for achievable scheme design, together with a corresponding low-complexity method for converse bounds.

\subsection{Learning-based caching work}
Existing learning-based caching techniques mainly focus on practical scenarios with unknown and time-varying content popularity, and are predominantly based on reinforcement learning (RL) or deep reinforcement learning (DRL). Several works \cite{zhang2020RLcaching,zhou2018RLcaching,sengupta2014RLcaching} estimate content popularity online and dynamically update cache contents, often relying on coded placement strategies such as Maximum Distance Separable (MDS) coding. However, these approaches primarily optimize the cache placement phase and do not explicitly design coded multicast transmissions in the delivery phase.
Further studies also consider cooperative caching, content sharing, or scheduling among multiple base stations. Some works \cite{wu2021RLcooperativecaching,gao2020RLcooperativecaching} employ fractional caching by optimizing the proportions of each file cached across different nodes, while another work \cite{song2017RLsharingcahing} considers file-level caching decisions, determining whether each entire file should be cached. By jointly optimizing cache placement, inter-base-station content sharing, and content acquisition costs, these approaches improve cache hit rates and reduce overall network costs. However, this line of work also does not address the explicit construction of coded multicast transmissions in coded caching systems.

The work \cite{zhang2019RLdoublecaching} combines learning mechanisms with coded caching in dynamic network environments to jointly optimize placement and delivery strategies. In this approach, RL is used to make high-level decisions, such as strategy selection or scheduling, while the underlying coded caching mechanisms are treated as fixed. 
Another work \cite{naderializadeh2019RLdeliverycaching} applies RL directly to the delivery phase, with the objective of learning efficient coded transmission strategies under a given cache placement. In this setting, the cache placement is assumed to be known and arbitrary at the bit level, and the RL determines how XOR combinations are formed. However, the resulting multicast coding decisions are typically learned in an unstructured manner, without explicitly exploiting the combinatorial structures.
RL-based approaches, such as \cite{shan2025TIMRL}, are not well suited to this setting, as they typically rely on repeated interactions with a fixed environment. In contrast, the MACC delivery problem involves dynamically varying conflict graphs induced by different user demands and access topology, which significantly limits the generalization of environment-specific RL policies and leads to high training complexity.

\subsection{Contributions}
This paper investigates the MACC system with arbitrary user-cache access topology. By reformulating the MACC delivery problem within a universal graph-based framework, we develop efficient delivery schemes and practical performance benchmarks for large-scale and irregular MACC systems. The main contributions of this work are summarized as follows.

\begin{itemize}
    \item We propose a universal graph-based formulation for the MACC delivery problem with arbitrary user-cache access topology. Under this formulation, decoding conflicts among requested packets induce a conflict graph, and the coded multicast delivery problem is reduced to a graph coloring problem, where each color corresponds to a multicast transmission.

    \item Based on the conflict graph formulation, we first apply the classical DSatur greedy graph coloring algorithm to construct coded multicast transmissions. We show that the DSatur algorithm achieves transmission loads close to the IC converse bound, thereby serving as a strong and practical benchmark. Motivated by its performance but high computational complexity, we further develop a scalable and unsupervised learning-based graph coloring framework using GNNs, which achieves transmission loads close to the IC converse bound with significantly lower computational complexity than the DSatur algorithm. In addition, the proposed scheme also has good generalization performance across different access topologies and  user numbers.

    \item We extend the IC converse bound for uncoded cache placement in \cite{wan2020index} to MACC systems with arbitrary access topology. To overcome the prohibitive complexity of converse bound evaluation in large-scale MACC systems, we further propose a low-complexity greedy approximation that closely matches the IC converse bound.
   
    \item Extensive numerical results with 10 cache-nodes and 4 to 20 users validate the effectiveness of the proposed framework. 
    Specifically, the DSatur algorithm is shown to achieve transmission loads very close to the IC converse bound, with a maximum gap no larger than $1.1\%$. 
    Building on this benchmark, the proposed GNN-based approach achieves transmission loads with a slight increase
    of up to 10\% compared to the DSatur algorithm, while reducing the computational runtime by a factor of $2$ to $30$ as the number of users increases. 
    Compared with the Potts-based method in~\cite{schuetz2022potts}, the proposed approach achieves substantial improvements in both computational complexity and coloring performance. In particular, the runtime is reduced by roughly several hundred times, while the number of colors required by the Potts-based method is consistently at least $1.2$ times that of our approach. 
    In addition, the proposed greedy converse bound closely approximates the IC converse bound, achieving more than $97\%$ of the bound while requiring only $10^{-2}$ to $10^{-6}$ of the computational runtime as the number of users grows.

\end{itemize}

\subsection{Paper organization}
The rest of this paper is organized as follows. Section II introduces the MACC system model with arbitrary user-cache access topology and presents a graph-based formulation that transforms the MACC delivery problem into a graph coloring problem. Section III provides the necessary preliminaries, including graph coloring approaches and the IC converse bound. Section IV introduces the proposed AI-aided delivery framework for MACC systems based on the conflict graph. Section V presents a low-complexity greedy approximation of the IC converse bound. Simulation results and performance evaluations are provided in Section VI. Finally, Section~VII concludes the paper.

\subsection{Notations}
The following notations are used throughout this paper unless otherwise stated.
\begin{itemize}
    \item For a set $\mathcal{A}$, $|\mathcal{A}|$ denotes its cardinality.
    \item For positive integers $a$, $b$, and $t$ with $a \le b$ and $t \le b$, we define $[a:b] \triangleq \{a, a+1, \ldots, b\}$. Moreover, $\binom{[b]}{t} \triangleq \{ \mathcal{A} \subseteq [b] \mid |\mathcal{A}| = t \}$ denotes the collection of all $t$-element subsets of $[b]$.
    \item For any $F \times K$ array $\mathbf{P}$, the entry in the $f^{\text{th}}$ row and $k^{\text{th}}$ column is denoted by $\mathbf{P}(f,k)$, where $f \in [F]$ and $k \in [K]$.
    \item The matrix $[{\bf A}; {\bf B}]$ is expressed in a format similar to Matlab, equivalent to $\begin{bmatrix}
     {\bf A}\\ 
     {\bf B}
 \end{bmatrix} $.
\end{itemize}

\section{System Model}
In this section, we first introduce the MACC system with arbitrary user-cache access topology. Then, we present a graph-theoretic formulation of the MACC system.

\subsection{Multi-access coded caching}\label{sec-macc}
\begin{figure}[htbp]
    \centering
    \includegraphics[width=0.8\textwidth]{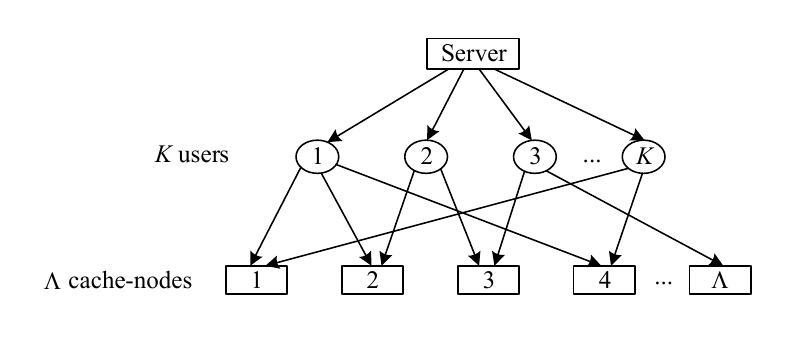}
    \caption{MACC system model with arbitrary user-cache access topology.}
    \label{fig-model}
\end{figure}

We consider the MACC system model \cite{hachem2017multi-level}, illustrated in Fig.~\ref{fig-model}, where a server has access to a library $\mathcal{W}=\{W_1,W_2, \ldots, W_N\}$ of $N$ equal-size files and is connected to $K$ users via an error-free broadcast link. The system contains $\Lambda$ cache-nodes, each of size $M$ files. Unlike the shared-link setting, in which each user is associated with a dedicated cache-node, in the MACC model, each user $k\in[K]$ can access an arbitrary subset of cache-nodes, denoted by $\mathcal{A}_k \subseteq [\Lambda]$. We refer to the collection $\mathcal{A} = \{\mathcal{A}_k \mid k \in [K]\}$ as the user-cache access topology.

A MACC scheme with an arbitrary user-cache access topology $\mathcal{A}$ consists of two phases.
\begin{itemize}
    \item \textbf{Placement phase:} Each file is divided into $F$ packets of equal size. Without knowledge of future requests, each cache-node $\lambda \in [\Lambda]$ stores a subset of packets from each file, subject to its storage constraint $M$. The content stored at cache-node $\lambda$ is denoted by $\mathcal{Z}_\lambda$. Each user $k \in [K]$ can retrieve all packets stored at the cache-nodes in its accessible set $\mathcal{A}_k$, i.e., the content accessible to user $k$ is $\mathcal{Z}'_k = \bigcup_{\lambda \in \mathcal{A}_k} \mathcal{Z}_\lambda$.
    
    \item \textbf{Delivery phase:} Each user randomly requests one file. Let $\mathbf{d} = (d_1, d_2, \ldots, d_K)$ denote the request vector. Based on $\mathbf{d}$, the cached contents $\mathcal{Z}_\lambda$, and the access topology $\mathcal{A}$, the server broadcasts $S_{\mathbf{d}}$ coded packets such that each user can recover its requested file using the received transmissions along with its accessible cache content $\mathcal{Z}'_k$.
\end{itemize}

In this system, the goal is to minimize 
 the worst-case load of the broadcast downlink (from the server to users), given by
\begin{equation}\label{eq-R}
    R = \max_{\mathbf{d} \in [N]^K} \frac{S_{\mathbf{d}}}{F},
\end{equation}
i.e., the maximum (over the user demands) length of the transmitted server message $S_{\mathbf{d}}$, normalized by the length of one file $F$ (in packets).


For the coded caching problem in the MACC system, a combinatorial framework called the MACC Placement Delivery Array (MACC-PDA) was proposed in \cite{cheng2021maccpda} to systematically characterize MACC schemes.

\begin{definition}[MACC-PDA, \cite{cheng2021maccpda}] \label{def-maccpda} 
In a MACC system, the placement and delivery phases can be jointly represented using a structured array called a MACC-PDA. It is composed of the node-placement array $\mathbf{C}$, user-retrieve array $\mathbf{U}$ and user-delivery array $\mathbf{Q}$.
\begin{itemize}
    \item \textbf{Node-placement array $\mathbf{C}$}. An $F\times \Lambda$ node-placement array $\mathbf{C}$ consists of stars and nulls, where $F$ and $\Lambda$ represent the number of packets of each file and the number of cache-nodes, respectively. For any $f\in[F]$ and $\lambda\in[\Lambda]$, the entry $\mathbf{C}(f,\lambda)=*$ if and only if the $\lambda^{\text{th}}$ cache-node stores the $f^{\text{th}}$ packet of each file $W_n$. 
    \item \textbf{User-retrieve array $\mathbf{U}$}.
    An $F \times K$ user-retrieve array $\mathbf{U}$ consists of stars and nulls, where $F$ and $K$ represent the subpacketization of each file and the number of users, respectively. For any $f\in[F]$ and $k\in[K]$, the entry $\mathbf{U}(f,k)=*$ if and only if the $f^{\text{th}}$ packet of each file $W_n$ is stored in at least one cache-node belonging to the user-cache access set $\mathcal{A}_k$. 
    \item \textbf{User-delivery array $\mathbf{Q}$}. An $F\times K$ user-delivery array $\mathbf{Q}$ consists of stars and integers in $[S]$, where the stars in $\mathbf{Q}$ have the same meaning as those in $\mathbf{U}$. Here, $S$ denotes the total number of multicast messages transmitted by the server during the delivery phase, an integer entry $s\in[S]$ indicates that the corresponding packet is
    delivered via the $s^{\text{th}}$ multicast transmission.
\end{itemize}
\hfill
$\square$ 
\end{definition}

It has been shown in \cite{cheng2021maccpda} that if the user-delivery array $\mathbf{Q}$ in Definition~\ref{def-maccpda} satisfies the following Condition, each user can successfully recover its requested file using its retrievable cache content and  received multicast packets.

\textit{Condition 1 \cite{yan2017placement}:  For any two distinct entries $\mathbf{Q}(f_1,k_1)$ and $\mathbf{Q}(f_2,k_2)$, $\mathbf{Q}(f_1,k_1)=\mathbf{Q}(f_2,k_2)=s$ is an integer only if
    \begin{itemize}
        \item [a.] $f_1\neq f_2$, $k_1\neq k_2$, i.e., they lie in distinct rows and columns; and
        \item [b.] $\mathbf{Q}(f_1,k_2)=\mathbf{Q}(f_2,k_1)=*$, i.e., the corresponding $2\times  2$ subarray formed by rows $f_1$, $f_2$ and columns $k_1$, $k_2$ must be of the following form
        \begin{equation}
        \left(\begin{array}{cc}
            s & * \\
            * & s
        \end{array}
        \right) \quad \text{or} \quad
        \left(\begin{array}{cc}
            * & s \\
            s & *
        \end{array}
        \right). 
        \end{equation}
    \end{itemize}}

\begin{example}\label{ex-maccpda}
Consider a MACC system with $K=5$ users and $\Lambda=4$ cache-nodes, as illustrated in Fig.~\ref{fig-ex-top}. The transformation procedure from the node-placement array $\mathbf{C}$ to the user-retrieve array $\mathbf{U}$ and the user-delivery array $\mathbf{Q}$ is shown in Fig.~\ref{fig-ex-macc-CUQ}. The user-cache access topology is given by $\mathcal{A}=\{\mathcal{A}_k \mid k \in [5]\}$, where
\begin{equation}\label{eq-maccpda-top}
\mathcal{A}_1 = \{1,2\}, \quad
\mathcal{A}_2 = \{1,3\}, \quad
\mathcal{A}_3 = \{4\}, \quad
\mathcal{A}_4 = \{2\}, \quad
\mathcal{A}_5 = \{3\}.
\end{equation}
Each cache-node stores the following subsets of packets according to the array $\mathbf{P}$ in Fig. \ref{fig-ex-macc-CUQ},
\begin{equation}
\begin{aligned}
\mathcal{Z}_1 &= \{W_{n,1},W_{n,2},W_{n,3} \mid n \in [6]\}, &
\mathcal{Z}_2 &= \{W_{n,1},W_{n,4},W_{n,5} \mid n \in [6]\}, \\
\mathcal{Z}_3 &= \{W_{n,2},W_{n,4},W_{n,6} \mid n \in [6]\}, &
\mathcal{Z}_4 &= \{W_{n,3},W_{n,5},W_{n,6} \mid n \in [6]\}.
\end{aligned}
\end{equation}
The corresponding node-placement array $\mathbf{C}$ is obtained by extracting all star entries from the MN PDA $\mathbf{P}$.
The user-retrieve array $\mathbf{U}$ is then constructed from $\mathbf{C}$ according to the user-cache access topology in \eqref{eq-maccpda-top}. Specifically, the $k^{\text{th}}$ column of $\mathbf{U}$ is obtained by taking the union of the star entries in the columns of $\mathbf{C}$ indexed by $\mathcal{A}_k$, i.e., $\mathbf{U}(f,k)=*$ if there exists some $\lambda \in \mathcal{A}_k$ such that $\mathbf{C}(f,\lambda)=*$. For example, the star entries in columns~$1$ and~$2$ of $\mathbf{P}$ are given by
\begin{equation}
\mathbf{P}(1,1)=\mathbf{P}(2,1)=\mathbf{P}(3,1)=*,\mathbf{P}(1,2)=\mathbf{P}(4,2)=\mathbf{P}(5,2)=*,
\end{equation} 
so we have
\begin{equation}
\mathbf{C}(1,1)=\mathbf{C}(2,1)=\mathbf{C}(3,1)=*,\mathbf{C}(1,2)=\mathbf{C}(4,2)=\mathbf{C}(5,2)=*,
\end{equation}
and  user~$1$ can access cache-nodes $\mathcal{A}_1=\{1,2\}$. 
As a result, the first column of $\mathbf{U}$ satisfies
\begin{equation}
    \mathbf{U}(1,1)=\mathbf{U}(2,1)=\mathbf{U}(3,1)=\mathbf{U}(4,1)=\mathbf{U}(5,1)=*.
\end{equation}
Finally, the user-delivery array $\mathbf{Q}$ is constructed by filling the null entries of $\mathbf{U}$ with integers such that Condition 1 is satisfied.
\hfill $\square$ 

\begin{figure}
\centering
\includegraphics[width=0.8\textwidth]{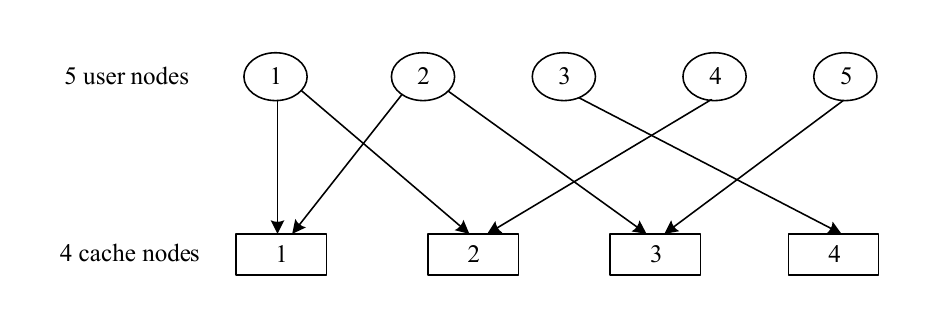}
\caption{User-cache access topology with $5$ users and $4$ cache-nodes.}
\label{fig-ex-top}
\end{figure}

\begin{figure}
\centering
\includegraphics[width=1\textwidth]{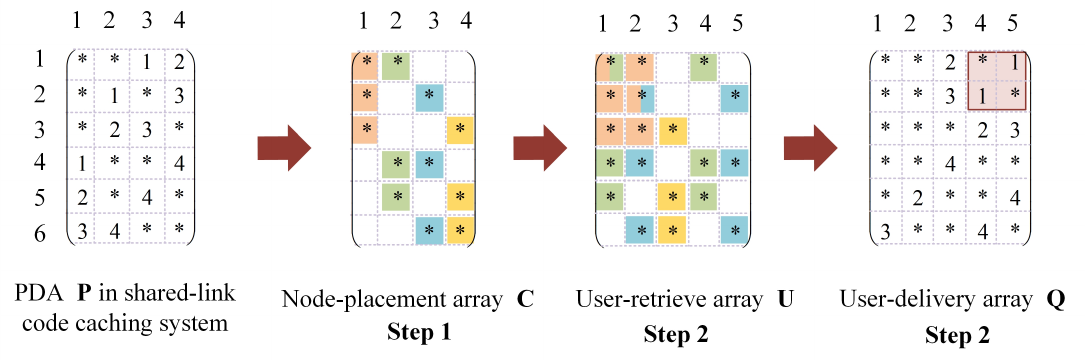}
\caption{Transformation from the MN PDA $\mathbf{P}$ to a MACC scheme via arrays $\mathbf{C}$, $\mathbf{U}$, and $\mathbf{Q}$.}
\label{fig-ex-macc-CUQ}
\end{figure}
\end{example}

However, existing schemes such as \cite{cheng2021maccpda, cheng2025t_designmacc, brunero2022fundamentalmacc} are primarily tailored to specific user-cache access topology. To the best of our knowledge, a universal coded caching scheme that can accommodate MACC systems with arbitrary access topology is still lacking. This observation motivates us to investigate a more universal design framework. In the following, we show that the user-delivery array $\mathbf{Q}$, constructed based on the user-retrieve array $\mathbf{U}$ in Definition~\ref{def-maccpda}, can be reformulated as a graph coloring problem. This reformulation constitutes the foundation of our proposed approach. 

\subsection{Transformation from MACC-PDA to a graph coloring problem}\label{sec-MACC-graph coloring}
We establish a graph-theoretic reformulation of the problem. Specifically, we show that designing the user-delivery array $\mathbf{Q}$ in Definition~\ref{def-maccpda}, which characterizes the multi-access coded caching scheme, is equivalent to a graph coloring problem. This equivalence provides an efficient perspective and enables the utilization of well-established graph coloring methods. We first introduce the concept of graph coloring for undirected graphs, which is essential for the transformation from the MACC system to a graph coloring problem.

\begin{definition}[Undirected graph and graph coloring~\cite{graphbook}]\label{def-graph-coloring}
An undirected graph is an ordered pair $\mathcal{G}=(\mathcal{V},\mathcal{E})$, where $\mathcal{V}$ is a finite and nonempty set of vertices and 
$\mathcal{E} \subseteq \{\{u,v\}\mid u,v\in\mathcal{V},\, u\neq v\}$ is a set of edges connecting distinct vertices; vertices $u$ and $v$ are said to be adjacent if $\{u,v\}\in\mathcal{E}$.  
A \emph{proper coloring} of $\mathcal{G}$ assigns colors to vertices such that no two adjacent vertices share the same color. 
If $k$ colors are used, the coloring is called a \emph{$k$-coloring}. 
The minimum such $k$ is the \emph{chromatic number} $\chi(\mathcal{G})$.
\hfill $\square$ 
\end{definition}

To design the user-delivery array $\mathbf{Q}$ from the user-retrieve array $\mathbf{U}$, the key challenge lies in assigning integer entries to the null positions of $\mathbf{U}$ such that Condition 1 is satisfied. To this end, we construct a conflict graph $\mathcal{G}=(\mathcal{V},\mathcal{E})$, which is an undirected graph as defined in Definition~\ref{def-graph-coloring}. In this graph, each vertex corresponds to one null entry in $\mathbf{U}$, and an edge is placed between two vertices if their corresponding entries violate Condition 1. Consequently, non-adjacent vertices can be assigned the same color.
In the resulting graph $\mathcal{G}=(\mathcal{V},\mathcal{E})$, a \emph{proper coloring} guarantees that adjacent vertices (i.e., conflicting entries) are assigned different colors, while non-adjacent vertices may share the same color. Therefore, finding a \emph{proper coloring} of $\mathcal{G}$ yields a valid assignment of integers to the null entries of $\mathbf{U}$ that satisfies Condition 1. Since each color corresponds to one multicast transmission (i.e., one integer in the user-delivery array $\mathbf{Q}$), minimizing the number of colors is equivalent to minimizing the transmission load defined in~\eqref{eq-R}. 
Hence, our objective is to determine the \emph{chromatic number} of the graph $\mathcal{G}$.

In the following, we present an example to illustrate the transformation from the user-retrieve array $\mathbf{U}$ to the user-delivery array $\mathbf{Q}$.

\begin{example}\label{ex-procedure}
In this example, we consider a $3\times 3$ user-retrieve array $\mathbf{U}$ shown in Fig.~\ref{fig-ex-procedure}(a), which consists of both star and null entries. The complete transformation procedure from $\mathbf{U}$ to the user-delivery array $\mathbf{Q}$ is illustrated in Fig.~\ref{fig-ex-procedure}, which consists of the following $4$ steps.
\begin{figure}
\centering
\includegraphics[width=1\textwidth]{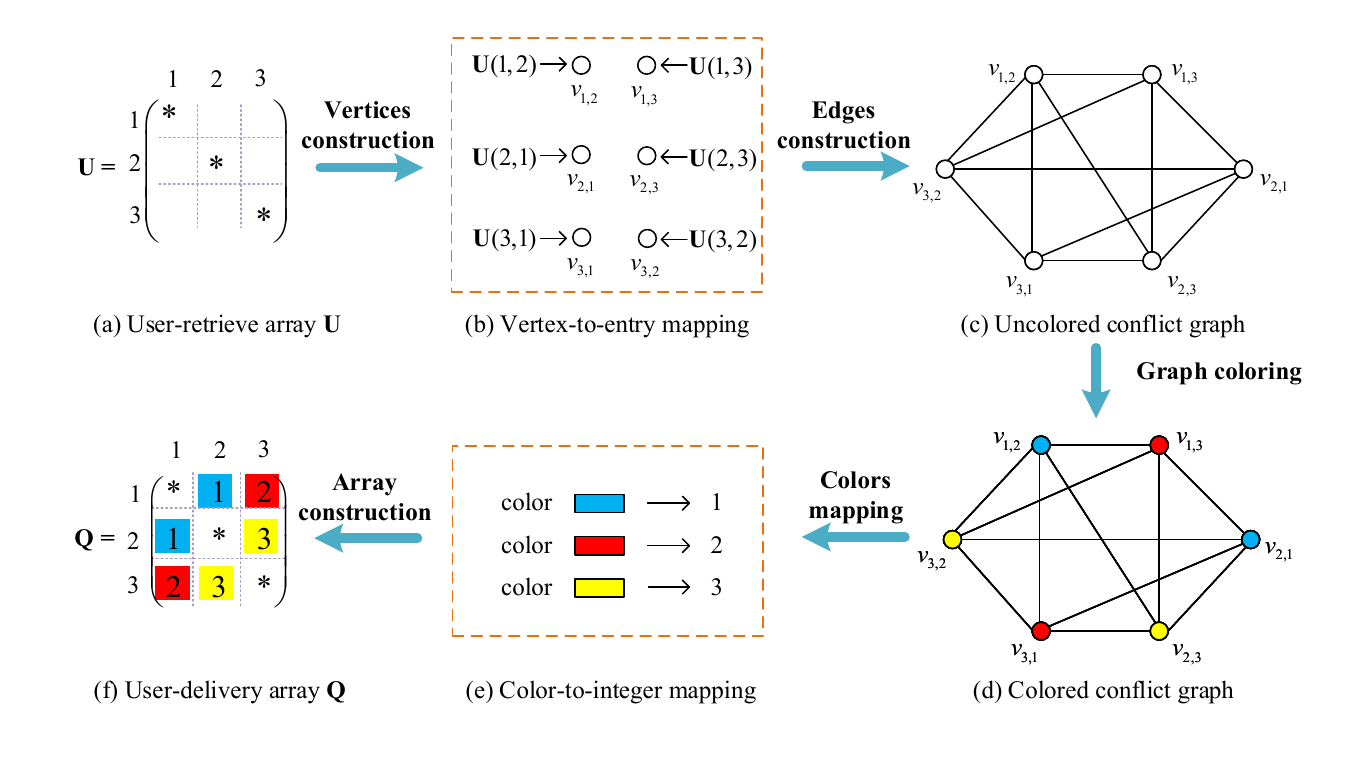}
\caption{Transformation from  user-retrieve array $\mathbf{U}$ to user-delivery array $\mathbf{Q}$ in Example~\ref{ex-procedure}.}
\label{fig-ex-procedure}
\end{figure}

\begin{itemize}

\item \textbf{Vertex construction}:
Each null entry $\mathbf{U}(f,k)$ in the user-retrieve array $\mathbf{U}$ is mapped to a unique vertex $v_{f,k}$ in the undirected graph. Accordingly, the vertex set $\mathcal{V}$ of graph $\mathcal{G}$ consists of all vertices corresponding to null entries in $\mathbf{U}$.
In Fig.~\ref{fig-ex-procedure}(a), there are six null entries, namely $\mathbf{U}(1,2)$, $\mathbf{U}(1,3)$, $\mathbf{U}(2,1)$, $\mathbf{U}(2,3)$, $\mathbf{U}(3,1)$, and $\mathbf{U}(3,2)$. Therefore, the vertex set is $\mathcal{V}=\{v_{1,2},v_{1,3},v_{2,1},v_{2,3},v_{3,1},v_{3,2}\}$, as illustrated in Fig.~\ref{fig-ex-procedure}(b).

\item \textbf{Edge construction}:  
Two vertices $v_{f_1,k_1}$ and $v_{f_2,k_2}$ are connected by an edge $\{v_{f_1,k_1}, v_{f_2,k_2}\}\in\mathcal{E}$ if and only if the corresponding entries $\mathbf{U}(f_1,k_1)$ and $\mathbf{U}(f_2,k_2)$ cannot simultaneously satisfy Condition 1. Thus, the edge set $\mathcal{E}$ consists of all pairs of vertices whose corresponding null entries are in conflict.
For instance, for the null entry $\mathbf{U}(1,2)$, only its pairing with $\mathbf{U}(2,1)$ satisfies Condition 1, whereas all other pairings violate the condition. Hence, vertex $v_{1,2}$ is connected to vertices $v_{1,3}$, $v_{2,3}$, $v_{3,1}$, and $v_{3,2}$. Applying the same rule to all vertices yields the set of edges listed in~\eqref{eq-ex-edge}. 
 \begin{equation}\label{eq-ex-edge}
    \begin{aligned}
        v_{1,2}:& \{v_{1,2},v_{1,3}\}, \{v_{1,2},v_{2,3}\}, \{v_{1,2},v_{3,1}\}, \{v_{1,2},v_{3,2}\};\\
        v_{1,3}:& \{v_{1,3},v_{1,2}\}, \{v_{1,3},v_{2,1}\}, \{v_{1,3},v_{2,3}\}, \{v_{1,3},v_{3,2}\};\\
        v_{2,1}:& \{v_{2,1},v_{1,3}\}, \{v_{2,1},v_{2,3}\}, \{v_{2,1},v_{3,1}\}, \{v_{2,1},v_{3,2}\};\\
        v_{2,3}:& \{v_{2,3},v_{1,2}\}, \{v_{2,3},v_{1,3}\}, \{v_{2,3},v_{2,1}\}, \{v_{2,3},v_{3,1}\};\\
        v_{3,1}:& \{v_{3,1},v_{1,2}\}, \{v_{3,1},v_{2,1}\}, \{v_{3,1},v_{2,3}\}, \{v_{3,1},v_{3,2}\};\\
        v_{3,2}:& \{v_{3,2},v_{1,2}\}, \{v_{3,2},v_{1,3}\}, \{v_{3,2},v_{2,1}\}, \{v_{3,2},v_{3,1}\}.
    \end{aligned}
    \end{equation}  
Since the graph is undirected, each edge is unordered, and the complete edge set $\mathcal{E}$ is given by
\begin{equation}
\begin{aligned}
\mathcal{E} = \{& \{v_{1,2},v_{1,3}\}, \{v_{1,2},v_{2,3}\}, \{v_{1,2},v_{3,1}\},\{v_{1,2},v_{3,2}\}, \{v_{1,3},v_{2,1}\}, \{v_{1,3},v_{2,3}\},\\
&\{v_{1,3},v_{3,2}\}, \{v_{2,1},v_{2,3}\}, \{v_{2,1},v_{3,1}\}, \{v_{2,1},v_{3,2}\}, \{v_{2,3},v_{3,1}\}, \{v_{3,2},v_{3,1}\} \}.
\end{aligned}
\end{equation}

\item \textbf{Graph coloring and integer mapping}:  
Following Definition~\ref{def-graph-coloring}, colors are assigned to the vertices of $\mathcal{G}$ such that the adjacent vertices receive different colors, while non-adjacent vertices may share the same color. By construction, non-adjacent vertices correspond to entries that already satisfy Condition 1.
Vertices $v_{1,2}$ and $v_{2,1}$ are non-adjacent and are therefore assigned the same color. Proceeding similarly for all vertices and mapping colors to integers yields
\begin{equation}\label{eq-ex-Coloring}
\begin{aligned}
    C(v_{1,2})=C(v_{2,1})=1,\quad
    C(v_{1,3})= C(v_{3,1})=2,\quad
    C(v_{2,3})=C(v_{3,2})=3.
\end{aligned}
\end{equation}

\item \textbf{Array construction}:  
Finally, according to the vertex index $(f,k)$ and its assigned color $s$, the color $s$ is placed in the corresponding entry $\mathbf{U}(f,k)$, producing the user-delivery array $\mathbf{Q}$. Specifically, in this example, $\mathbf{Q}(1,2)=\mathbf{Q}(2,1)=1$, $\mathbf{Q}(1,3)=\mathbf{Q}(3,1)=2$, and $\mathbf{Q}(2,3)=\mathbf{Q}(3,2)=3$.
As shown in~\cite{yan2017placement}, the resulting array $\mathbf{Q}$ achieves the minimum transmission load in~\eqref{eq-R}. Equivalently, the graph $\mathcal{G}$ admits a \emph{proper coloring} with chromatic number $\chi(\mathcal{G})=3$.
\end{itemize}
\hfill $\square$ 
\end{example}


Based on the above transformation, the construction of the user-delivery array $\mathbf{Q}$ can be equivalently interpreted as a graph coloring problem on the conflict graph $\mathcal{G}=(\mathcal{V},\mathcal{E})$. Consequently, the MACC delivery design problem reduces to finding a \emph{proper coloring} of $\mathcal{G}$ using as few colors as possible, i.e., a \emph{$k$-coloring} with a minimum $k$. However, determining the exact \emph{chromatic number} of an arbitrary graph $\mathcal{G}$ is a well-known NP-hard combinatorial optimization problem in graph theory \cite{1982NP-hard}. As a result, computing the optimal user-delivery array $\mathbf{Q}$ via exact graph coloring becomes computationally prohibitive for large-scale MACC systems.


\subsection{Main objective of the paper}\label{Objective}
The objective of this paper is to develop a universal and low-complexity 
delivery scheme with the MN cache placement at the cache-nodes,\footnote{
The special case where $K > \Lambda$, and each user is connected with a single cache is isomorphic to the decentralized caching scheme where there are Lambda cache types and each user loads one of these types at random and independently of the other users. This case has been widely investigated either as a tool to reduce subpacketization (especially in connection with MISO-delivery), as in \cite{lampiris2018adding}
or as a tool to construct fully decentralized caching schemes where the cache construction does not depend on the number of users in the system  (e.g.,\cite{bayat2021coded}). In all these cases, the cache types are MN construction replacing the number of users $K$ with the number of cache-nodes $\Lambda$.

}
for arbitrary user-cache access topology.

In addition, with the MN cache placement, we also aim to establish a universal and low-complexity converse bound on the transmission load for arbitrary user-cache access topology.

\section{Preliminaries}
In this section, we review the graph coloring approaches \cite{brelaz1979dsatur, goudet2022aitaub, li2020rethinking, schuetz2022potts, zhang2024gin}. Then we extend the IC converse bound on the transmission load developed in~\cite{wan2020index} to the MACC system considered in this paper.

\subsection{Graph coloring approaches}\label{sec-graph-coloring}
A wide range of graph coloring algorithms has been studied in the literature, including the local search-based methods such as the tabu search~\cite{hertz1987tabu} and the evolutionary algorithms  \cite{islam2021evolutionary}. While these methods can achieve good performance in certain settings, they are often sensitive to algorithmic parameters and incur substantial computational overhead for large-scale or structurally complex graphs.
Greedy heuristics are generally more efficient. Among them, the DSatur algorithm~\cite{brelaz1979dsatur} is a widely used greedy coloring method that dynamically selects vertices according to their saturation degree ($SD$), defined as the number of distinct colors assigned to neighboring vertices. \emph{The main idea of the DSatur algorithm is to prioritize vertices that are most constrained by previously assigned colors, thereby reducing the risk of future color conflicts.}
By prioritizing the most constrained vertices, the DSatur algorithm produces colorings close to the IC converse bound in the MACC delivery problem (as shown in Section~\ref{sec-experiments}). 
However, its worst-case time complexity is on the order of $\mathcal{O}(|\mathcal{V}|^2 + |\mathcal{E}|)$, and the number of vertices in the conflict graph can scale as $|\mathcal{V}|=\mathcal{O}(K^{t+1})$ under MN uncoded placement. Moreover, decoding constraints in MACC systems often induce dense conflict graphs, further exacerbating the computational burden. These limitations motivate the development of low-complexity alternatives.

Recently, GNNs \cite{kipf2016GNN} have emerged as a promising learning-based approach for graph coloring. By leveraging message-passing mechanisms, GNNs can learn expressive vertex embeddings that capture both local neighborhood information and global graph structure. Existing GNN-based graph coloring methods, however, exhibit notable limitations. Some works focus on single-graph optimization \cite{goudet2022aitaub} and lack cross graph generalization. 
Other methods consider relaxed or indirect formulations of the graph coloring problem. Examples include Graph Discrimination Networks (GDNs)~\cite{li2022GDNcoloring}, Minimal Cost Graph Neural Networks (MCGNNs)~\cite{gao2024gnn}, and Graph Isomorphism Network (GIN)-based approaches~\cite{zhang2024gin}. These methods learn representations related to coloring constraints but do not explicitly guarantee conflict-free color assignments.

    Potts-based approaches~\cite{schuetz2022potts} formulate graph coloring as an energy minimization problem, where conflicts are penalized in the objective. However, the obtained solutions may still contain residual conflicts, and the authors apply iterative postprocessing to eliminate them, which incurs significant computational overhead. The work in~\cite{ijaz2022colornum}, on the other hand, focuses on predicting the \emph{chromatic number} rather than constructing a \emph{proper coloring}.

Overall, existing graph coloring methods, both classical and learning-based, face a fundamental tradeoff among solution quality, generalization capability, and computational complexity. These limitations motivate the development of alternative GNN-based formulations that can efficiently produce high-quality colorings while generalizing across diverse MACC conflict graphs.

\subsection{Converse bound based on index coding}\label{sec-converse-IC}
The converse bound under uncoded cache placement developed in~\cite{wan2020index} establishes a fundamental connection between coded caching and index coding (IC). Specifically, when the cache placement phase is uncoded and the cache contents are fixed, the delivery phase of a caching system can be viewed as equivalent to an index coding problem.
In the considered MACC setting, for the given MN cache-node placement and user-cache access topology $\mathcal{A}$, the cache contents available to each user can be uniquely determined. Therefore, the framework of IC converse bound  in~\cite{wan2020index} can be directly applied to derive a lower bound on the transmission load of the MACC system.
The work in~\cite{wan2020index} shows that, for any given cache placement and user permutation, an acyclic set of subfiles can be constructed to yield a linear inequality on the transmission load, expressed as the sum of the lengths of the subfiles in the selected acyclic set. The overall converse bound is then obtained by aggregating such inequalities over all possible user permutations. 

\begin{lemma}[Converse bound~\cite{wan2020index}]\label{le-converse}
Consider a MACC system with $K$ users, $\Lambda$ cache-nodes, an arbitrary user-cache access topology and a symmetric uncoded cache placement.\footnote{\label{foot:symmetric}Let    $W_{n,\mathcal{J}}$ denote the packets of $W_n$ which are stored exclusively by   cache-nodes in $\mathcal{J} \subseteq [\Lambda]$, where $|\mathcal{J}|=t$.  If  an uncoded cache placement is symmetric across the files, all files are partitioned and cached identically, and 
 thus $|W_{1,\mathcal{J}}|=\ldots=|W_{N,\mathcal{J}}|$; thus we can omit the file index and denote the number of packets exclusively stored  by cache-nodes in $\mathcal{J}$ as $|W_{\mathcal{J}}|$ for any file.} 
 Let $\mathcal{A}_k \subseteq [\Lambda]$ denote the set of cache-nodes accessible to user $k \in [K]$. For any user permutation $\mathbf{u} = (u_1, \dots, u_K)$, the transmission load satisfies
\begin{equation}\label{eq-le-R_u}
R_{\mathbf{u}} \ge 
\sum_{i\in[K]}
\sum_{\mathcal{J}_{u_i} \subseteq [\Lambda] \setminus \bigcup_{j=1}^i \mathcal{A}_{u_j} }
\frac{|W_{\mathcal{J}_{u_i}}|}{F},
\end{equation}
The worst-case transmission load under the symmetric uncoded cache placement is bounded by
\begin{equation}\label{eq-le-R_max}
R \ge 
\max_{\mathbf{u}\in\mathfrak{U}}
R_{\mathbf{u}}, 
\end{equation}
where $\mathfrak{U}$ denotes the set of all user permutations.
\hfill $\square$ 
\end{lemma}
The inequality in~\eqref{eq-le-R_u} corresponds to a single acyclic-set inequality associated with a specific user permutation $\mathbf{u}$, and follows directly from the IC converse bound by constructing an acyclic set according to a permutation order $\mathbf{u}$. 
 By taking the maximum of~\eqref{eq-le-R_u} over all user permutations $\mathbf{u}$, we obtain a converse bound on the worst-case transmission load of the MACC system in ~\eqref{eq-le-R_max}.

Evaluating the converse bound in Lemma~\ref{le-converse} requires enumerating all user permutations (i.e., $K!$ permutations). Consequently, the number of inequalities grows factorially with the number of users $K$, resulting in  a computational complexity on the order of
\begin{equation}
\mathcal{O}\left( |\mathfrak{U}| \right) =\mathcal{O} \left( |\text{Perm}(K,K)| \right)=\mathcal{O} \left( K! \right).
\end{equation}
For example, when $N=K=5$, there are $|\mathfrak{U}| = 5! = 120$ distinct inequalities, and this number increases exponentially as $K$ grows. 


\section{AI-Aided Framework for MACC Systems with Arbitrary Topology}\label{sec-GNN-train}

Designing the delivery phase for MACC systems with arbitrary user-cache access topology is challenging. As shown in Section~\ref{sec-graph-coloring}, this problem can be reformulated as graph coloring on the induced conflict graph. However, classical heuristics such as the DSatur algorithm become computationally expensive for large MACC graphs. This motivates a learning-based approach that can produce low-complexity color predictions and generalize across diverse user-cache access topologies. 
Moreover, many learning-based graph coloring methods are trained and evaluated on graphs with relatively regular structures. In contrast, the conflict graphs of the considered MACC systems, induced by MACC-PDA constructions and  user-cache access topologies, exhibit highly variable structures that are difficult to characterize in advance. This variability presents additional challenges for generalization and necessitates a framework that can effectively handle diverse graph structures. Therefore, we propose the following AI-aided framework.


 Overall, the above fact motivates us to design a new universal AI-aided coloring framework for the MACC systems with arbitrary  user-cache access topology, which has a significantly lower complexity than the existing schemes and achieves a performance approaching the converse bound.

\begin{figure}
    \centering
    \includegraphics[width=1\textwidth]{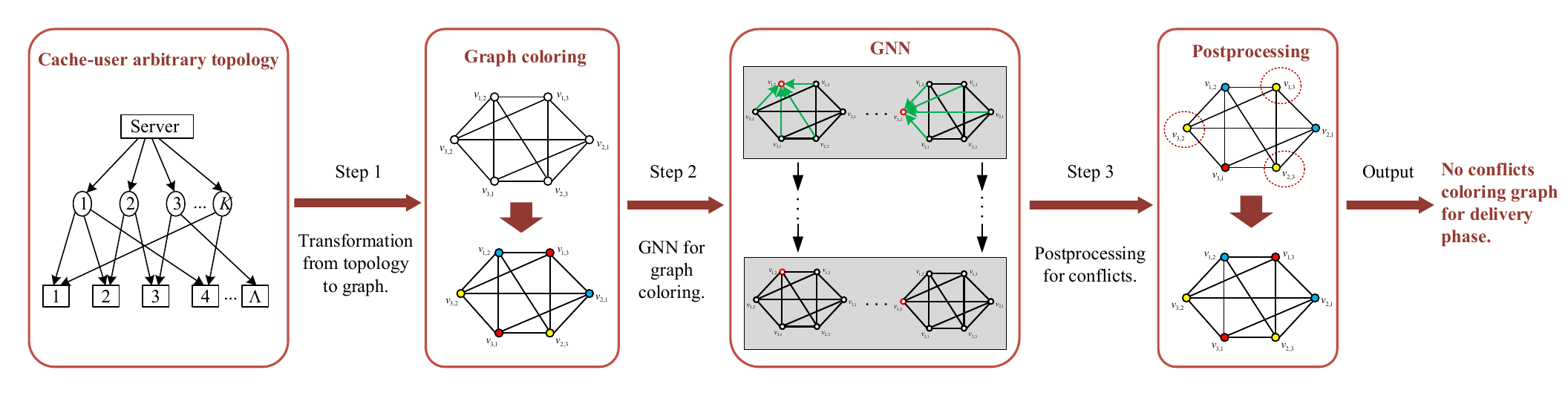}
    \caption{AI-aided framework for the delivery phase in MACC systems with arbitrary user-cache access topology.}
    \label{fig-framework}
\end{figure}
\textit{Overview of the proposed framework:} We propose an AI-aided framework to optimize the delivery phase in MACC systems with the MN cache placement and arbitrary user-cache access topology. 
With a user-cache access topology as the input and the resulting delivery scheme as the output, the proposed framework based on unsupervised learning contains three main steps, as illustrated in Fig. \ref{fig-framework}.
Step~1, detailed in Section~\ref{sec-MACC-graph coloring}, transforms the MACC-PDA problem into an equivalent conflict graph formulation. Step 2 employs a GNN to predict color assignments for the vertices of the conflict graph. Step 3 applies a lightweight postprocessing procedure to ensure that the predicted coloring corresponds to a feasible delivery scheme. By combining learning-based predictions with structured postprocessing, the framework achieves low-complexity delivery design with strong generalization across  different user-cache topologies and   different numbers of users.
  
\subsection{Step 1: Transformation from MACC-PDA to graph coloring}

This step takes the user-cache access topology $\mathcal{A}$ as input and constructs the corresponding undirected conflict graph $\mathcal{G}$ as output. The detailed transformation procedure was already provided in Section~\ref{sec-MACC-graph coloring}.

\subsection{Step 2: GNN-Based color prediction}
In this step, the conflict graph $\mathcal{G}$ obtained from Step 1 serves as the input of GNNs to predict the color assignments for each vertex. The output is a colored graph, which directly corresponds to the delivery scheme of the MACC system.
Although the learning-based coloring approach does not require highly expressive or global graph-level structural representations, it focuses on local conflicts and ensuring color assignments satisfy decoding constraints. This lightweight design allows the GNN to focus on relevant local information, significantly reducing computational complexity while achieving near-optimal performance. The neighbor-enhanced message-passing mechanism emphasizes relevant neighbors, and the embedding-based color projection relaxes discrete coloring into a continuous space, facilitating learning and enabling efficient postprocessing.

The step consists of the following components.

\subsubsection{Input representation and preprocessing}
The input to this module is the undirected conflict graph constructed from the MACC delivery problem. For each vertex $v$, we use its degree $D(v)$ as a scalar input feature, denoted by $x_v=D(v)$. Stacking all vertex features gives the feature matrix $\mathbf{X}=[x_1;x_2;\dots;x_{|\mathcal{V}|}]\in\mathbb{R}^{|\mathcal{V}|\times 1}$. To ensure consistent scaling across different graphs, $\mathbf{X}$ is standardized using the global mean and standard deviation computed from the training set. The graph structure is encoded by the edge index $\mathbf{e}\in\mathbb{Z}^{2\times |\mathcal{E}|}$ in the PyTorch Geometric sparse format. Accordingly, each graph is represented as a PyTorch Geometric data object $\text{Data}(x=\mathbf{X}, \text{edge\_index}=\mathbf{e})$, which serves as the input to the subsequent GNN model. 
Since graph coloring labels are permutation-invariant, explicit color labels are not used as supervision. 
The output dimension of the GNN is specified by a predefined maximum color budget $C_{\max}$, which represents the number of candidate color classes in the prediction layer and is fixed before training. 
During inference, the trained GNN directly predicts color assignments under this fixed budget, without requiring any graph coloring algorithm to be run beforehand. The setting of $C_{\max}$ will be discussed in Section~\ref{sec-experiments}. 

\subsubsection{Model Architecture}\label{sec-GNN-arch}
\begin{figure}[t]
    \centering
    \includegraphics[width=1\textwidth]{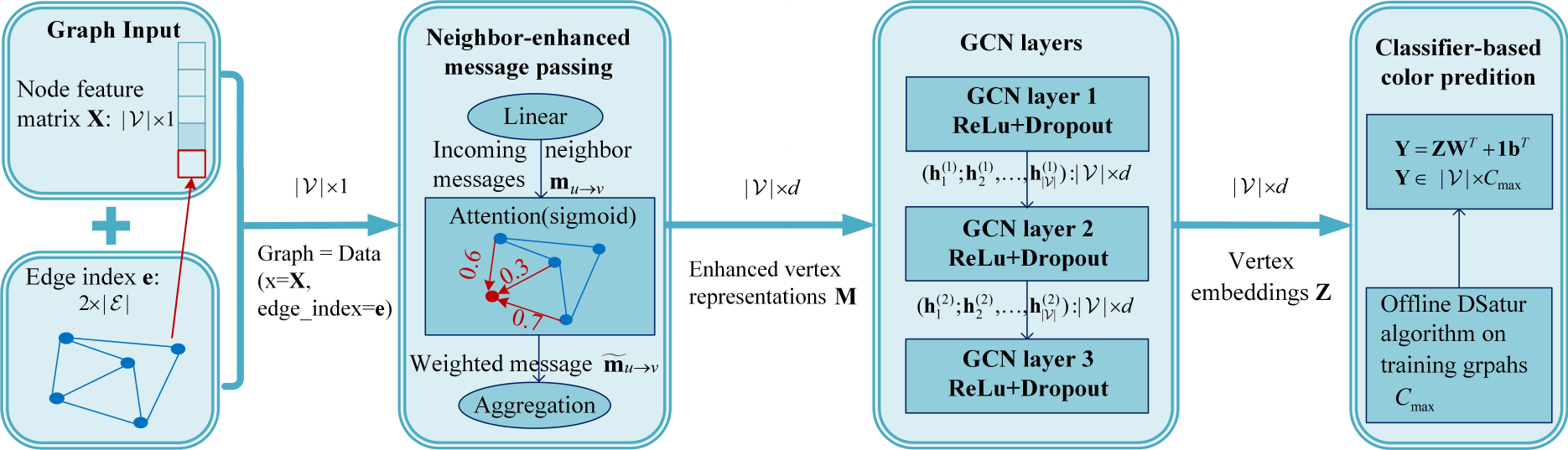}
    \caption{Architecture of the proposed GNN model for graph coloring in MACC systems.}
    \label{fig-model-arch}
\end{figure}

Our GNN model follows a standard message-passing framework and is tailored to learn proper colorings for conflict graphs arising from MACC systems with arbitrary user-cache access topology. The network takes as input the vertex feature matrix $\mathbf{X}\in\mathbb{R}^{|\mathcal{V}|\times 1}$ and the edge index $\mathbf{e}\in\mathbb{Z}^{2\times|\mathcal{E}|}$ of the preprocessed conflict graph. It first extracts structure-aware vertex representations through a neighbor-enhanced message-passing module and stacked GCN layers, and then maps the resulting embeddings to color logits through a classifier head under a fixed color budget $C_{\max}$. The output logits are transformed into soft color assignments by a softmax function and further converted into a discrete coloring for the delivery phase. Fig.~\ref{fig-model-arch} shows the overall architecture.

\textbf{Neighbor-enhanced message passing:}  
This module takes as input the vertex feature matrix $\mathbf{X}\in\mathbb{R}^{|\mathcal{V}|\times 1}$ and edge index $\mathbf{e}\in\mathbb{Z}^{2\times |\mathcal{E}|}$ of the preprocessed conflict graph, and outputs an enhanced vertex representation matrix $\mathbf{M}\in\mathbb{R}^{|\mathcal{V}|\times d}$. Although the conflict graph is undirected, message passing is implemented over directed edge pairs $u \to v$ for information aggregation.

Specifically, for each edge $u \to v$, the scalar feature $x_u$ of the source vertex $u$ is first transformed into a $d$-dimensional row message vector through a shared linear mapping, i.e., $\mathbf{m}_{u\to v} = x_u \mathbf{w}_m \in \mathbb{R}^{1\times d}$,
where $\mathbf{w}_m \in \mathbb{R}^{1\times d}$ is a trainable vector shared across all vertices and edges. The resulting message $\mathbf{m}_{u\to v}$ encodes the information of vertex $u$ in a form suitable for subsequent weighting.
Next, a scalar importance weight is computed as
\begin{equation}
    s_{u\to v} = \sigma(\mathbf{m}_{u\to v}\mathbf{a}) \in (0,1),
\end{equation}
where $\mathbf{a} \in \mathbb{R}^{d\times 1}$ is a trainable vector and $\sigma(\cdot)$ denotes the sigmoid function.\footnote{We use the sigmoid function instead of softmax so that each neighbor message is weighted independently. This avoids diluting the contribution of individual edges for high-degree vertices and leads to more stable training.} The weighted message is then given by $\tilde{\mathbf{m}}_{u\to v} = s_{u\to v}\mathbf{m}_{u\to v} \in \mathbb{R}^{1\times d}$.
Finally, the updated representation of vertex $v$ is obtained by aggregating the weighted messages from all its neighbors, i.e.,
\begin{equation}
    \mathbf{m}_v = \sum_{u \in \mathcal{N}(v)} \tilde{\mathbf{m}}_{u\to v} \in \mathbb{R}^{1\times d}.
\end{equation}
Stacking the updated representations of all vertices gives the output matrix $\mathbf{M}=[\mathbf{m}_1;\mathbf{m}_2;\dots;\mathbf{m}_{|\mathcal{V}|}] \in \mathbb{R}^{|\mathcal{V}|\times d}$, which is then passed to the subsequent GCN layers.

\textbf{Graph Convolutional Network (GCN) layers:}  
The enhanced vertex representation matrix $\mathbf{M}\in\mathbb{R}^{|\mathcal{V}|\times d}$ produced by the neighbor-enhanced message-passing module is then fed into stacked GCN layers to propagate information over the conflict graph and capture both local and higher-order structural dependencies. Let $\mathbf{H}^{(0)}=\mathbf{M}$ denote the input to the GCN backbone. For each layer $l$, let $\mathbf{H}^{(l)}\in\mathbb{R}^{|\mathcal{V}|\times d}$ be the vertex embedding matrix, whose $v^\text{th}$ row is denoted by $\mathbf{h}_v^{(l)}\in\mathbb{R}^{1\times d}$. At the $l^\text{th}$ layer, the input consists of the vertex embeddings $\mathbf{h}_v^{(l-1)}$ for all $v\in\mathcal{V}$ together with the graph connectivity encoded by $\mathcal{N}(v)$, and the output is the updated embedding $\mathbf{h}_v^{(l)}\in\mathbb{R}^{1\times d}$.
Formally, the update for vertex $v$ at layer $l$ is given by
\begin{equation}
\mathbf{h}_v^{(l)}=\sigma\!\left(
\sum_{u\in\mathcal{N}(v)\cup\{v\}}
\frac{1}{\sqrt{\hat{D}(v)\hat{D}(u)}}
\mathbf{h}_u^{(l-1)}\mathbf{W}^{(l)}
\right),
\end{equation}
where $\hat{D}(v)$ denotes the degree of vertex $v$ after adding self-loops, $\mathbf{W}^{(l)}\in\mathbb{R}^{d\times d}$ is a trainable weight matrix, and $\sigma(\cdot)$ denotes a nonlinear activation function. The symmetric normalization term $\frac{1}{\sqrt{\hat{D}(v)\hat{D}(u)}}$ prevents high-degree vertices from dominating the aggregation. In our implementation, three GCN layers with ReLU activations and dropout are employed. By stacking these layers, each vertex progressively aggregates information from increasingly larger neighborhoods in the conflict graph. The final GCN output is denoted by $\mathbf{H}^{(L)}\in\mathbb{R}^{|\mathcal{V}|\times d}$, which is then passed to a linear projection layer and the subsequent classifier head for color prediction.

\textbf{Classifier-based color prediction:}  
This stage maps the final vertex embeddings obtained from the GCN backbone to color logits under a fixed color budget. The input to this module is the vertex embedding matrix $\mathbf{Z}=[\mathbf{z}_1;\mathbf{z}_2;\ldots;\mathbf{z}_{|\mathcal{V}|}]\in\mathbb{R}^{|\mathcal{V}|\times d}$, where each row $\mathbf{z}_v\in\mathbb{R}^{1\times d}$ denotes the embedding of vertex $v$ after the linear projection layer. A linear classifier then maps these embeddings to the output logit matrix $\mathbf{Y}=\mathbf{Z}\mathbf{W}^T+\mathbf{1}\mathbf{b}^T$, where $\mathbf{W}\in\mathbb{R}^{C_{\max}\times d}$ and $\mathbf{b}\in\mathbb{R}^{C_{\max}\times 1}$ are trainable parameters, $\mathbf{1}\in\mathbb{R}^{|\mathcal{V}|\times 1}$ is an all-ones vector, and $\mathbf{Y}\in\mathbb{R}^{|\mathcal{V}|\times C_{\max}}$ is the output logit matrix.
For each vertex $v$, the corresponding row $\mathbf{y}_v\in\mathbb{R}^{1\times C_{\max}}$ of $\mathbf{Y}$ provides the predicted logits over the candidate colors. Applying a softmax operation to $\mathbf{y}_v$ yields a differentiable soft color assignment for vertex $v$. These soft assignments are then converted into a discrete coloring for use in the MACC delivery phase.

Overall, by integrating the neighbor-enhanced message-passing module, stacked GCN layers, and the classifier-based color prediction module, the proposed architecture learns expressive vertex embeddings and produces high-quality graph colorings, making it well-suited for efficient delivery design in MACC systems.

\subsubsection{GNN Training}
The proposed GNN model is trained in an unsupervised manner to learn \emph{proper colorings} of the conflict graphs arising from the MACC delivery phase with arbitrary user-cache access topology. Supervised training with explicit color assignments from the DSatur algorithm is unsuitable for this problem, since graph coloring labels are arbitrary up to permutation: any permutation of color indices corresponds to an equivalent valid coloring as long as adjacent vertices receive different colors. Consequently, a supervised loss such as cross-entropy would incorrectly penalize solutions that differ only by a relabeling of colors.

To overcome this limitation, we adopt a Potts-inspired unsupervised training strategy~\cite{schuetz2022potts}. For a graph $\mathcal{G}=(\mathcal{V},\mathcal{E})$, let $C(v)$ denote the discrete color assigned to vertex $v$. The corresponding Potts Hamiltonian can be written as
\begin{equation}
H_{\text{Potts}} = J \sum_{\{u,v\} \in \mathcal{E}} \delta\big(C(u)=C(v)\big),
\end{equation}
where $J>0$ is the interaction strength and $\delta(\cdot)$ is the indicator function. This formulation assigns positive energy to color conflicts between adjacent vertices.

To enable gradient-based optimization, we relax the discrete color assignments to continuous soft assignments. Specifically, for each vertex $v$, the classifier output row $\mathbf{y}_v\in\mathbb{R}^{1\times C_{\max}}$ is transformed into a soft color assignment
\begin{equation}
\mathbf{p}_v=\mathrm{Softmax}(\mathbf{y}_v)\in[0,1]^{1\times C_{\max}}.
\end{equation}
Based on this relaxation, we define the Potts-inspired loss as
\begin{equation}
\mathcal{L}_{\rm Potts}(\theta)
=
\sum_{\{u,v\}\in\mathcal{E}} \mathbf{p}_u \mathbf{p}_v^T,
\end{equation}
which penalizes color overlap between adjacent vertices. Minimizing this loss reduces the probability that neighboring vertices are assigned the same color. Moreover, the loss is permutation-invariant with respect to color labels, since any permutation of the color indices yields the same objective value. Furthermore, the formulation is inherently scalable, as the loss is defined over local edge-wise interactions, resulting in a computational complexity that scales linearly with the number of edges. Moreover, the independent structure of these interactions enables fully parallelizable optimization across vertices, making it efficient for large-scale graphs and large color spaces.

To further encourage compact color usage, we introduce a color-count regularization term. For each graph $\mathcal{G}$ in a mini-batch $\mathcal{G}_B$, let $\widetilde{C}_{\mathcal{G}}$ denote the soft number of activated colors predicted by the GNN, and let $C_{\mathcal{G}}^{\star}$ denote the reference color number obtained from the DSatur algorithm. The color-count loss is defined as
\begin{equation}
\mathcal{L}_{\rm color\text{-}count}(\theta)
=
\sum_{\mathcal{G}\in \mathcal{G}_B}
\left(
\widetilde{C}_{\mathcal{G}}-C_{\mathcal{G}}^{\star}
\right)^2.
\end{equation}
The overall training objective is given by
\begin{equation}\label{eq-loss}
\mathcal{L}_{\rm train}(\theta)
=
\mathcal{L}_{\rm Potts}(\theta)
+
\beta \mathcal{L}_{\rm color\text{-}count}(\theta),
\end{equation}
where $\beta>0$ controls the tradeoff between conflict reduction and color-count matching. The color-count regularization is used only during training to guide the optimization toward compact color usage. During inference, no loss term is evaluated or optimized; the trained GNN directly produces an initial coloring under the fixed color budget $C_{\max}$, which is then refined by the postprocessing step. Therefore, neither the color-count regularization nor the DSatur algorithm is needed during testing.

\subsubsection{Training procedure} \label{sec-training-procedure}
The GNN parameters $\theta$ are optimized end-to-end using mini-batch stochastic gradient descent. In each training iteration, a mini-batch of conflict graphs is merged into a single disconnected graph using the batching operator provided by PyTorch Geometric, and the training loss is computed according to \eqref{eq-loss}. The output dimension of the model is fixed to the predefined color budget $C_{\max}$ determined from the training graphs. The parameters are updated using the AdamW optimizer together with a cosine annealing schedule for the learning rate.

During inference, a discrete coloring is obtained by assigning each vertex the color with the highest predicted probability
\begin{equation}
    C(v)=\arg\max_{c\in[C_{\max}]} p_{v,c},
\end{equation}
where $p_{v,c}$ denotes the predicted probability of assigning color $c$ to vertex $v$. Since the training objective penalizes coloring conflicts only through soft assignments, the argmax operation may still produce conflicting edges, i.e.,
\begin{equation}
\mathcal{E}_{\text{conf}}
=
\bigl\{\{u,v\}\in\mathcal{E}\mid C(u)=C(v)\bigr\}.
\end{equation}
Therefore, a postprocessing step is applied to repair the remaining conflicts and obtain a proper coloring for the graph.

\subsection{Step 3: Postprocessing}
\label{sec-postprocessing}
After training, the GNN produces an initial vertex coloring that may still contain conflicts, i.e., some adjacent vertices may be assigned the same color. To obtain a \emph{proper coloring}, we apply a postprocessing procedure, summarized in Algorithm~\ref{alg-postprogress}. This procedure employs the color selection strategy in Algorithm~\ref{alg-select} to efficiently assign colors while controlling the total number of colors used.

\begin{algorithm}[htbp]
\caption{Conflict repair procedure}
\label{alg-postprogress}
\begin{algorithmic}[1]
\State \textbf{Input:} Graph $\mathcal{G}=(\mathcal{V},\mathcal{E})$, initial coloring $C$
\State \textbf{Output:} A proper coloring $C$
\State Collect used colors $\mathcal{C}_{\text{used}} \gets \{ C(v) \mid v\in\mathcal{V} \}$
\State Identify conflict vertices $\mathcal{V}_{\text{conf}} \gets \{ v \mid \exists(u,v)\in\mathcal{E},\; C(u)=C(v)\}$
\If{$\mathcal{V}_{\text{conf}}\neq\emptyset$}
    \State Compute saturation degree of vertices in $\mathcal{V}_{\text{conf}}$, $SD(v)=|\{C(u)\mid u\in\mathcal{N}(v)\}|$
    \State $\mathbf{V}_{\text{conf}} \gets$ list of $\mathcal{V}_{\text{conf}}$ sorted in decreasing $SD(v)$
    \For{$v \in \mathbf{V}_{\text{conf}}$}
        \State $(C(v),\mathcal{C}_{\text{used}}) \gets \textsc{ColorSelect}(v, \mathcal{G}, C, \mathcal{C}_{\text{used}})$ in Algorithm \ref{alg-select}
    \EndFor
\EndIf
\State \Return $C$
\end{algorithmic}
\end{algorithm}

\begin{algorithm}[htbp]
\caption{Color selection}
\label{alg-select}
\begin{algorithmic}[1]
\Function{ColorSelect}{$v, \mathcal{G}, C, \mathcal{C}_{\text{used}}$}
    \State $\mathcal{C}_{\mathcal{N}(v)} \gets \{ C(u)\mid u\in\mathcal{N}(v)\}$
    \State Select the minimum available color $c_{\text{new}} \gets \min\{c\in\mathcal{C}_{\text{used}} \mid c\notin \mathcal{C}_{\mathcal{N}(v)}\}$ 

    \If{$c_{\text{new}}$ does not exist}
        \State Add a new color $c_{\text{new}} \gets \max(\mathcal{C}_{\text{used}})+1$
        \State Update used colors $\mathcal{C}_{\text{used}} \gets \mathcal{C}_{\text{used}} \cup \{c_{\text{new}}\}$
    \EndIf

    \State \Return $(c_{\text{new}}, \mathcal{C}_{\text{used}})$
\EndFunction
\end{algorithmic}
\end{algorithm}

\subsubsection{Conflict repair}
The conflict repair procedure takes as input the graph $\mathcal{G}=(\mathcal{V},\mathcal{E})$ and the initial vertex coloring $C$ produced by the proposed GNN. Lines 3-4 collect all colors currently used $\mathcal{C}_{\text{used}}$ in the graph, and identify the conflict vertices $\mathcal{V}_{\text{conf}}$ whose current color assignments violate the coloring constraints. Lines 5-7 compute the saturation degree $SD$ for each vertex $v\in \mathcal{V}_{\text{conf}}$ and sort them by decreasing order of $SD$ to obtain the ordered list $\mathbf{V}_{\text{conf}}$. Lines 8--10, process each vertex $v \in \mathbf{V}_{\text{conf}}$ using the color selection procedure in Algorithm~\ref{alg-select}, which assigns the smallest available color not used by neighboring vertices or introduces a new color if necessary, while updating $\mathcal{C}_{\text{used}}$. Since each processed vertex is assigned a color distinct from all its current neighbors, the repair step eliminates the conflicts incident to that vertex without introducing new conflicts. Therefore, after all vertices in $\mathbf{V}_{\text{conf}}$ are processed once, Line~12 returns the updated coloring as a \emph{proper coloring} $C$.


\subsubsection{Color selection}

The color selection procedure takes as input a conflict vertex $v$, the graph $\mathcal{G}$, the current coloring $C$, and the set of used colors $\mathcal{C}_{\text{used}}$. Line 2 collects the set of colors $\mathcal{C}_{\mathcal{N}(v)} = \{ C(u) \mid u \in \mathcal{N}(v) \}$ used by the neighbors of vertex $v$. Line 3 selects the smallest color in $\mathcal{C}_{\text{used}}$ that does not appear in $\mathcal{C}_{\mathcal{N}(v)}$, and assigns it as $c_{\text{new}}$. Lines 4-7 handle the case where no available color exists, a new color is added as $c_{\text{new}} =\max(\mathcal{C}_{\text{used}})+1$, and the set of used colors is updated according to $\mathcal{C}_{\text{used}} \gets \mathcal{C}_{\text{used}} \cup \{ c_{\text{new}} \}$. Finally, Line 8 returns the assigned color $c_{\text{new}}$ together with the updated set $\mathcal{C}_{\text{used}}$, ensuring that conflicts are resolved efficiently while limiting the total number of colors.

\begin{remark}
The conflict repair procedure in Algorithm~\ref{alg-postprogress} is guaranteed to terminate in a finite number of steps. 
Finite-step convergence requires two conditions: 
(i) each update eliminates at least one existing conflict without introducing new conflicts, and 
(ii) a \emph{proper coloring} exists for the conflict graph.

The proposed Algorithm~\ref{alg-postprogress} satisfies these conditions. 
At each step, every conflicting vertex $v$ is reassigned a color that is distinct from the colors of its neighbors in $\mathcal{N}(v)$, either by selecting an admissible existing color or introducing a new color if necessary. 
Thus, each update resolves the conflict at $v$ without creating additional conflicts. 
Moreover, since the conflict graph $\mathcal{G}=(\mathcal{V},\mathcal{E})$ contains a finite number of vertices, a \emph{proper coloring} always exists; in the worst case, assigning distinct colors to all vertices yields a feasible solution. Therefore, the total number of conflict resolutions is bounded, and Algorithm~\ref{alg-postprogress} converges to a \emph{proper coloring} in a finite number of updates.
\end{remark}

\begin{example}
We illustrate the conflict repair procedure with an  example. As shown in Fig.~\ref{fig-ex-postprocessing}, the initial coloring is given by
\begin{equation}
C=\{C(v_{1,2})=1, C(v_{1,3})=2, C(v_{2,1})=1, C(v_{2,3})=3, C(v_{3,1})=3, C(v_{3,2})=2\}.
\end{equation}
The set of used colors is $\mathcal{C}_{\text{used}}=\{1,2,3\}$, and the set of conflict vertices is $\mathcal{V}_{\text{conf}}=\{v_{1,3}, v_{2,1}, v_{3,2}\}$.
Since all vertices in $\mathcal{V}_{\text{conf}}$ have the same saturation degree $SD$, without loss of generality, the vertices are ordered according to their indices, yielding the ordered list $\mathbf{V}_{\text{conf}} = (v_{1,3}, v_{2,1}, v_{3,2})$.
The vertices in $\mathbf{V}_{\text{conf}}$ are then recolored sequentially using Algorithm~\ref{alg-select}. The neighboring color sets of the conflict vertices are
\begin{equation}
\mathcal{C}_{\mathcal{N}(v_{1,3})}=\{1,2\}, \quad
\mathcal{C}_{\mathcal{N}(v_{2,1})}=\{1,2,3\}, \quad
\mathcal{C}_{\mathcal{N}(v_{3,2})}=\{1,2,3\}.
\end{equation}
The vertex $v_{1,3}$ is assigned the minimum available color
\begin{equation}
c_{\text{new}} = \min \{ c \in \mathcal{C}_{\text{used}} \mid c \notin \mathcal{C}_{\mathcal{N}(v_{1,3})} \}
= \min \{ c \in \{1,2,3\} \mid c \notin \{1,2\} \} = 3.
\end{equation}
After updating the color of $v_{1,3}$, the remaining conflict vertices $v_{2,1}$ and $v_{3,2}$ are checked, and no further conflicts are detected in the graph. Consequently, a \emph{proper coloring} is obtained as
\begin{equation}
C=\{C(v_{1,2})=1,\; C(v_{1,3})=3,\; C(v_{2,1})=1,\; C(v_{2,3})=2,\; C(v_{3,1})=3,\; C(v_{3,2})=2\}.
\end{equation}

\begin{figure}[htbp]
\centering
\includegraphics[width=0.8\textwidth]{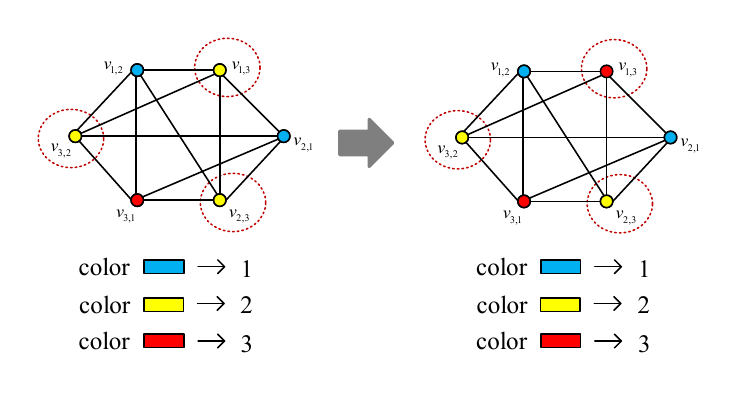}
\caption{The proposed postprocessing conflict repair procedure.}
\label{fig-ex-postprocessing}
\end{figure}

\hfill $\square$ 
\end{example}

In summary, the proposed AI-aided framework for MACC systems with arbitrary user-cache access topology integrates a GNN-based graph coloring predictor with a lightweight postprocessing conflict repair mechanism. The GNN captures the structural properties of MACC-induced conflict graphs and produces an initial coloring, while the postprocessing procedure resolves only the remaining conflicts, requiring less computation than the DSatur algorithm. As a result, the framework supports an efficient delivery phase and reduces the transmission load in MACC systems.

\section{Low-Complexity Converse Bound for MACC-PDA}\label{sec-greedy-converse}
Given a cache placement which is symmetric over files,   the IC converse bound presented in Lemma~\ref{le-converse} has  computational complexity that grows factorially with the number of users $K$, as it requires enumerating all user permutations, i.e., $\mathcal{O}(K!)$. Consequently, as $K$ grows, this exhaustive enumeration rapidly becomes infeasible, limiting the applicability of the IC converse bound to small-scale MACC systems.

To address this limitation, we develop a low-complexity converse bound that exploits the structural properties of the MACC-PDA and adopts a greedy selection procedure. The proposed bound substantially reduces the computational complexity while remaining highly accurate. In particular, it closely approximates the IC converse bound. 
Instead of exhaustively evaluating all permutations as in the IC converse bound, our method iteratively selects users according to the overlap among their inaccessible packet sets. At each iteration, the user whose inaccessible packet set has the largest intersection with the current accumulated set is selected. This greedy accumulation of intersections yields an efficient approximation of the IC converse bound while requiring only linear operations per iteration.
Given a cache placement that is symmetric over files, the procedure of the proposed greedy converse bound consists of three main steps.

\begin{itemize}

\item \textit{Extraction of inaccessible packets:} 
For each column of the user-retrieve array $\mathbf{U}$ in the MACC-PDA, we extract the set of packets that are not accessible to the corresponding users through their connected cache-nodes.
\item \textit{Greedy accumulation of intersections:} 
Initialize an empty set $\mathcal{S}$. At each iteration, among the remaining columns, select the column whose associated inaccessible packet set has the largest intersection with the current set $\mathcal{S}$. After selection, update $\mathcal{S}$ as the intersection with the selected set.

\item \textit{Approximate converse bound computation:} 
At each iteration, the size of the intersection between the newly selected set and the current set $\mathcal{S}$ is accumulated. According to \eqref{eq-R}, the sum of these intersection sizes yields the resulting transmission load, providing a relaxed approximation of the IC converse bound without enumerating all column permutations.

\end{itemize}

\begin{algorithm}
\caption{Algorithm for greedy converse bound  based on MACC-PDA}
\label{alg-greedy-converse} 
\begin{algorithmic}[1]
\State \textbf{Input:} User-retrieve array $\mathbf{U}$ of MACC-PDA
\State \textbf{Output:} Transmission load $R$
\State Initialize accumulated set $\mathcal{S} \gets \emptyset$, cumulative sum $S \gets 0$, remaining columns $\mathcal{K}_{\text{remain}} \gets [K]$, and inaccessible packet set of users $\mathcal{S}_k, k\in[K]$
\While{$\mathcal{K}_{\text{remain}} \neq \emptyset$}
    \For{each column $k \in \mathcal{K}_{\text{remain}}$}
        \If{$\mathcal{S} \neq \emptyset$}
        \State Compute intersection $\mathcal{I}_k \gets \mathcal{S} \cap \mathcal{S}_k$ 
        \Else 
        \State {$\mathcal{I}_k \gets \mathcal{S}_k $}
        \EndIf
    \EndFor
    \State Select column $k' \gets \arg\max_{k \in \mathcal{K}_{\text{remain}}} |\mathcal{I}_k |$ 
    \State Update accumulated set $\mathcal{S} \gets \mathcal{I}_{k'}$, cumulative sum $S \gets S + |\mathcal{S}|$
    \State Remove $k'$ from $\mathcal{K}_{\text{remain}}$
\EndWhile
\State \Return $R=\frac{S}{F}$ 
\end{algorithmic}
\end{algorithm}

Given a cache placement which is symmetric over files,    the IC converse bound of a MACC system with $K$ users described in Section \ref{sec-converse-IC} requires $\mathcal{O}(K!)$ operations, becoming computationally prohibitive as $K$ increases. 
In contrast, the proposed greedy converse algorithm performs only $K$ sequential greedy selections. At each iteration, the algorithm computes intersections over the remaining users, resulting in an overall complexity of $\mathcal{O}(K^2)$. 
This represents a dramatic reduction from factorial to polynomial complexity, making the greedy converse bound scalable for large-scale MACC-PDA systems. 

To further illustrate the operation of the greedy converse algorithm, we present a small example based on the MACC-PDA in Example~\ref{ex-maccpda}. The column selection process and the resulting approximate transmission load are summarized in Table~\ref{tab-selection-columns}.

\begin{example}\label{ex-greedy-converse}
We consider the $6 \times 4$ node-placement array $\mathbf{C}$ and the $6 \times 5$ user-retrieve array $\mathbf{U}$ of the MACC-PDA as shown in Fig.~\ref{fig-ex-converse}. For the array $\mathbf{C}$, each cache-node $\lambda\in[4]$ adopts the MN placement.
Specifically,
\begin{equation}\label{eq-node-place}
\begin{aligned}
\mathcal{Z}_1 &= \{W_{n,\{1,2\}}, W_{n,\{1,3\}}, W_{n,\{1,4\}} \,|\, n\in[4]\},
\quad \mathcal{Z}_2 = \{W_{n,\{1,2\}}, W_{n,\{2,3\}}, W_{n,\{2,4\}} \,|\, n\in[4]\},\\
\mathcal{Z}_3 &= \{W_{n,\{1,3\}}, W_{n,\{2,3\}}, W_{n,\{3,4\}} \,|\, n\in[4]\},
\quad \mathcal{Z}_4 = \{W_{n,\{1,4\}}, W_{n,\{2,4\}}, W_{n,\{3,4\}} \,|\, n\in[4]\}.
\end{aligned}  
\end{equation}
According to the user-cache access topology $\mathcal{A}=\{\mathcal{A}_1=\{1,2\},\mathcal{A}_2=\{1,3\},\mathcal{A}_3=\{4\},\mathcal{A}_4=\{2\},\mathcal{A}_5=\{3\}\}$ in Fig.~\ref{fig-ex-top}, the packets accessible to user $k\in[5]$ are given by $\mathcal{Z}'_k = \bigcup_{\lambda\in\mathcal{A}_k} \mathcal{Z}_\lambda$. Thus,
the sets of accessible packets for each user $k$ are then given by
\begin{equation}\label{eq-accessible-packet}
\begin{aligned}
 \mathcal{Z}'_1&=\mathcal{Z}_1\cup\mathcal{Z}_2=\{W_{n,\{1,2\}},W_{n,\{1,3\}},W_{n,\{1,4\}},W_{n,\{2,3\}},W_{n,\{2,4\}}\,|\, n\in[4]\},\\
 \mathcal{Z}'_2&=\mathcal{Z}_1\cup\mathcal{Z}_3=\{W_{n,\{1,2\}},W_{n,\{1,3\}},W_{n,\{1,4\}},W_{n,\{2,3\}},W_{n,\{3,4\}}\,|\, n\in[4]\},\\  
 \mathcal{Z}'_3&=\mathcal{Z}_4=\{W_{n,\{1,4\}},W_{n,\{2,4\}},W_{n,\{3,4\}}\,|\, n\in[4]\},\\ 
 \mathcal{Z}'_4&=\mathcal{Z}_2=\{W_{n,\{1,2\}},W_{n,\{2,3\}},W_{n,\{2,4\}}\,|\, n\in[4]\},\\ 
 \mathcal{Z}'_5&=\mathcal{Z}_3=\{W_{n,\{1,3\}},W_{n,\{2,3\}},W_{n,\{3,4\}}\,|\, n\in[4]\}.
\end{aligned}    
\end{equation}
The sets of inaccessible packets for each user $k$ are
\begin{equation}\label{eq-inaccessible-packet}
\begin{aligned}
\bar{\mathcal{Z}}_1 &= \{W_{n,\{3,4\}} \,|\, n\in[4]\},
\quad\bar{\mathcal{Z}}_2 = \{W_{n,\{2,4\}} \,|\, n\in[4]\},\\
\quad\bar{\mathcal{Z}}_3 &= \{W_{n,\{1,2\}}, W_{n,\{1,3\}}, W_{n,\{2,3\}} \,|\, n\in[4]\},\\
\bar{\mathcal{Z}}_4 &= \{W_{n,\{1,3\}}, W_{n,\{1,4\}}, W_{n,\{3,4\}} \,|\, n\in[4]\},\\
\bar{\mathcal{Z}}_5 &= \{W_{n,\{1,2\}}, W_{n,\{1,4\}}, W_{n,\{2,4\}} \,|\, n\in[4]\}.
\end{aligned}
\end{equation}
We denote by $\mathcal{S}_k$ the index set of packets that are not accessible to user $k$ in \eqref{eq-inaccessible-packet}. 
Initially, the remaining column set is $\mathcal{K}_{\mathrm{remain}}=\{1,2,3,4,5\}$, the accumulated set is $\mathcal{S}=\emptyset$, and the cumulative sum is $S=0$. 
Note that  $\mathcal{K}_{\mathrm{remain}}$ keeps track of the columns (users) that have not yet been selected in the greedy procedure, 
$\mathcal{S}$ represents the intersection of indices of inaccessible packets corresponding to the currently selected columns,
which is used to compute the incremental transmission load contributed by the next selected column,  
and $S$ accumulates the total number of transmissions (transmission load) as columns are added. 
For notational simplicity, a subset such as $\{1,2\}$ is written in compact form as $\{12\}$. Accordingly,
\begin{equation}
\begin{aligned}
\mathcal{S}_1 &= \{34\}, 
\quad \mathcal{S}_2 = \{24\}, 
\quad \mathcal{S}_3 = \{\{12\},\{13\},\{23\}\}, \\
\mathcal{S}_4 &= \{\{13\},\{14\},\{34\}\}, 
\quad \mathcal{S}_5 = \{\{12\},\{14\},\{24\}\}.
\end{aligned}
\end{equation}

\begin{figure}[htbp]
    \centering
    \includegraphics[width=0.95\textwidth]{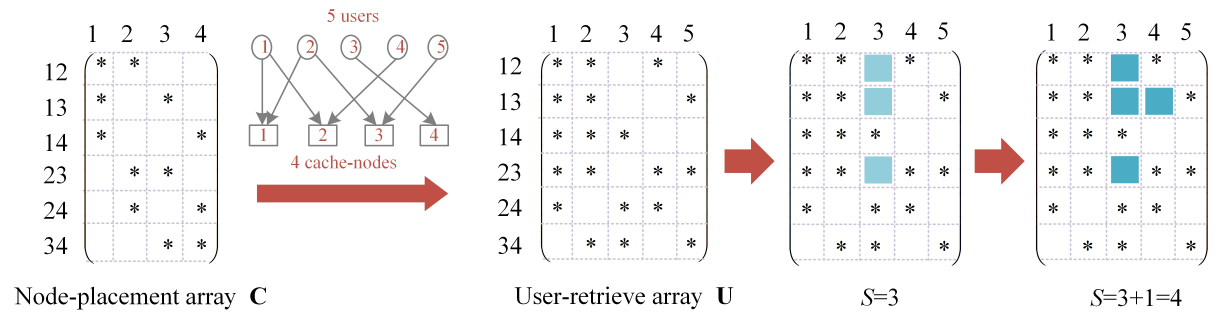}
    \caption{The procedure of the proposed greedy converse bound.}
    \label{fig-ex-converse}
\end{figure} 

By Lines 4-11 of Algorithm~\ref{alg-greedy-converse}, we compute the intersection $\mathcal{I}_k=\mathcal{S}\cap\mathcal{S}_k$ for each $k\in\mathcal{K}_{\text{remain}}=\{1,2,3,4,5\}$. Since the accumulated set is initially empty, i.e., $\mathcal{S}=\emptyset$, it follows that
\begin{equation}
\mathcal{I}_k=\mathcal{S}_k, k \in \{1,2,3,4,5\}.
\end{equation}
The corresponding cardinalities are
\begin{equation}
|\mathcal{I}_1|=|\mathcal{I}_2|=1,\quad
|\mathcal{I}_3|=|\mathcal{I}_4|=|\mathcal{I}_5|=3.
\end{equation}
Consequently, Line~12 selects a column with the largest intersection cardinality. Among columns ${3,4,5}$, which all attain the maximum value, without loss of generality, we select $k'=3$. Following Lines~13--15, the accumulated set is updated to $\mathcal{S}=\mathcal{I}_3$, the cumulative sum becomes $S=|\mathcal{S}|=3$, and column 3 is removed from $\mathcal{K}_{\text{remain}}$.
This greedy selection procedure is repeated until all columns have been processed. The detailed operations at each iteration are listed in Table~\ref{tab-selection-columns}. After the final iteration, Line~16 returns the greedy converse bound $R=\frac{S}{F}=\frac{4}{6}=\frac{2}{3}$.

We compare the proposed greedy converse bound with the IC converse bound in Section~\ref{sec-converse-IC} for the user-retrieve array $\mathbf{U}$. 
For the considered example, the files are partitioned according to the MN placement with parameter $t$, where each subfile $W_{\mathcal{J}}$ corresponds to a subset $\mathcal{J}\subseteq[\Lambda]$ with $|\mathcal{J}|=t$. 
Under this placement, the IC converse bound in \eqref{eq-le-R_max} can be written as
\begin{equation}\label{eq-ex-R_max}
\begin{aligned}
R &\ge \max_{\mathbf{u}\in\mathfrak{U}}\sum_{i\in[K]}\sum_{\substack{
\mathcal{J}_{u_i} \subseteq [\Lambda] \setminus \bigcup_{j=1}^i \mathcal{A}_{u_j} \\
|\mathcal{J}_{u_i}|=t}}\frac{|W_{\mathcal{J}_{u_i}}|}{F}.
\end{aligned}
\end{equation} 
Evaluating the IC converse bound requires examining all $K! = 5! = 120$ user permutations. After this exhaustive search on the user permutations, we find that the converse bound in~\eqref{eq-ex-R_max} becomes $R = 2/3$, which 
  exactly coincides with the greedy converse bound obtained in Example~\ref{ex-greedy-converse}. In contrast to the IC converse bound, which requires evaluating all $120$ permutations, the greedy approach only computes the intersections for the remaining columns at each iteration, with a total complexity order of $\mathcal{O}(K^2)$.
\begin{table}
\centering
\caption{Iteration steps of the greedy converse algorithm.}
\label{tab-selection-columns}
\begin{tabular}{|c|c|c|c|c|}
\hline
Step & $\mathcal{K}_{\text{remain}}$ & $\mathcal{I}_k$ & Chosen $k'$ & $\mathcal{S},|\mathcal{S}|,S$ \\
\hline
1 & $\{1,2,3,4,5\}$ &
\begin{tabular}{@{}c@{}}
$\mathcal{I}_1 = \{34\}$\\
$\mathcal{I}_2 = \{24\}$\\
$\mathcal{I}_3 = \{\{12\},\{13\},\{23\}\}$\\
$\mathcal{I}_4 = \{\{13\},\{14\},\{34\}\}$\\
$\mathcal{I}_5 = \{\{12\},\{14\},\{24\}\}$
\end{tabular} & $k'=3$ & \begin{tabular}{@{}c@{}}
 $\mathcal{S} = \{\{12\},\{13\},\{23\}\}$,\\ $|\mathcal{S}|=3,S=3$ 
\end{tabular}\\
\hline
2 & $\{1,2,4,5\}$ &
\begin{tabular}{@{}c@{}}
$\mathcal{I}_1=\mathcal{S}\cap\mathcal{S}_1=\emptyset$\\
$\mathcal{I}_2=\mathcal{S}\cap\mathcal{S}_2=\emptyset$\\
$\mathcal{I}_4 = \mathcal{S}\cap\mathcal{S}_4 = \{13\}$\\
$\mathcal{I}_5 = \mathcal{S}\cap\mathcal{S}_5 = \{12\}$\\
\end{tabular} & $k'=4$ & 
\begin{tabular}{@{}c@{}}
$\mathcal{S} = \{13\}, |\mathcal{S}|=1$\\
$S=3+1=4$ 
\end{tabular}\\
\hline
3 & $\{1,2,5\}$ &
\begin{tabular}{@{}c@{}}
$\mathcal{I}_1=\mathcal{S}\cap\mathcal{S}_1=\emptyset$\\
$\mathcal{I}_2=\mathcal{S}\cap\mathcal{S}_2=\emptyset$\\
$\mathcal{I}_5=\mathcal{S}\cap\mathcal{S}_5=\emptyset$\\
\end{tabular}
& $k'=1$ & 
$\mathcal{S}=\mathcal{I}_1, |\mathcal{S}|=0, S=4$ \\
\hline
4 & $\{2,5\}$ &
\begin{tabular}{@{}c@{}}
$\mathcal{I}_2=\mathcal{S}\cap\mathcal{S}_2=\emptyset$\\
$\mathcal{I}_5=\mathcal{S}\cap\mathcal{S}_5=\emptyset$
\end{tabular}
& $k'=2$ &
$\mathcal{S}=\mathcal{I}_2, |\mathcal{S}|=0,\ S=4$ \\
\hline
5 & $\{5\}$ &
\begin{tabular}{@{}c@{}}
$\mathcal{I}_5=\mathcal{S}\cap\mathcal{S}_5=\emptyset$
\end{tabular}
& $k'=5$ &
$\mathcal{S}=\mathcal{I}_5, |\mathcal{S}|=0,\ S=4$ \\
\hline
\end{tabular}
\end{table}
\hfill $\square$ 
\end{example}

\section{Experiments}\label{sec-experiments}

In this section, we evaluate the proposed AI-aided delivery framework for MACC systems with arbitrary user-cache access topology. Our experiments focus on the delivery-phase graph coloring performance of the proposed GNN-based method, and the accuracy and computational complexity of the proposed greedy converse bound. For the delivery phase, we compare the proposed GNN-based method with the DSatur algorithm and a Potts-based baseline~\cite{schuetz2022potts} in terms of the number of colors and runtime under different user numbers and memory ratios.

\subsection{Dataset generation} \label{sec-dataset}

To evaluate the proposed framework under the general MACC setting, we construct a dataset of conflict graphs induced by different user-cache access topologies.
Consider a MACC system with $K$ users and $\Lambda$ cache-nodes.
In practical edge caching systems, the number of cache-nodes is typically fixed due to deployment at predetermined locations, whereas the number of users may vary dynamically. Accordingly, in our experiments, we fix $\Lambda = 10$ and vary $K$ to evaluate the generalization of the proposed framework.

Each topology is specified by an access family $\mathcal{A}=\{\mathcal{A}_k \mid k\in[K]\}$, where $\mathcal{A}_k\subseteq[\Lambda]$ denotes the set of cache-nodes accessible to user $k$. All generated topologies satisfy the following structural constraints:
\begin{itemize}
    \item \textit{User coverage:} each user connects to at least one cache-node, i.e., $\mathcal{A}_k \neq \emptyset$ for all $k\in[K]$;
    \item \textit{Cache coverage:} each cache-node is accessed by at least one user.
\end{itemize}
The resulting access topologies exhibit heterogeneous and irregular connectivity patterns, including sparse configurations, partially overlapping access sets, and highly coupled user-cache connections. In particular, the access degrees $|\mathcal{A}_k|$ vary across users and topology instances, which induces conflict graphs with diverse sizes, degree distributions, and chromatic characteristics.

Under the MN placement, each user-cache access topology induces a corresponding MACC-PDA, which is further transformed into a conflict graph as described in Section~\ref{sec-MACC-graph coloring}. For each topology instance, we consider memory ratios $M/N\in\{0.1,0.2,\ldots,0.9\}$, yielding nine conflict graph instances per topology and increasing the diversity of the dataset.

The training, validation, and test sets are generated independently so that the topologies in the test set are unseen during training. The GNN is trained on a combined dataset containing all user numbers $K\in[4,20]$, with the number of cache-nodes fixed at $\Lambda=10$. For each user number $K$, we generate 2550 topology instances for training and validation, which are split with a ratio of $90\%$ and $10\%$, respectively. Each topology instance is combined with memory ratios $M/N\in\{0.1,0.2,\ldots,0.9\}$ to generate the corresponding conflict graph instances. For testing, we independently generate 50 unseen topology instances for each user number $K$, and each topology is evaluated under all memory ratios $M/N\in\{0.1,0.2,\ldots,0.9\}$.

\subsection{Implementation details}
The overall training and inference procedure follows Section~\ref{sec-training-procedure}, and the detailed GNN architecture is described in Section~\ref{sec-GNN-arch}. 
The GNN is trained on a combined dataset containing all user numbers $K \in [4,20]$ with the cache-node number fixed at $\Lambda=10$; see Section~\ref{sec-dataset} for details. 
All experiments use the hyperparameter settings summarized in Table~\ref{tab-training-config}. 

The model consists of three graph convolutional layers with 128 hidden dimensions and outputs vertex-level color logits, which are converted into color probabilities via a softmax operator. 
The output dimension of the prediction layer is specified by a maximum color budget $C_{\max}$. 
In our implementation, $C_{\max}$ is determined only from the graphs in the training set. 
Specifically, after generating the training set, we run the DSatur algorithm on each training conflict graph and set $C_{\max}=\max_{\mathcal{G}\in\mathcal{D}_{\mathrm{train}}} S_{\mathrm{DSatur}}(\mathcal{G})$, where $S_{\mathrm{DSatur}}(\mathcal{G})$ denotes the number of colors used by the DSatur algorithm on graph $\mathcal{G}$, and $\mathcal{D}_{\mathrm{train}}$ denotes the training set
\footnote{
The DSatur algorithm is used only to obtain a compact empirical color budget from the training graphs. 
Although $\Delta+1$ is a valid coloring upper bound, where $\Delta=\max_{v\in\mathcal{V}}D(v)$ denotes the maximum degree of the conflict graph, it is usually loose and would unnecessarily enlarge the GNN output dimension. 
Computing the minimum number of colors required for a proper coloring is computationally prohibitive for large conflict graphs, whereas the DSatur algorithm provides a computationally efficient heuristic with near-IC converse coloring performance in Fig. \ref{fig-DSvsIC}. 
Therefore, its coloring number provides a practical choice for $C_{\max}$.
}.
This procedure is used only to determine the fixed output dimension before training. 
During inference, $C_{\max}$ is kept fixed, and no graph coloring algorithm is run on the test graphs before GNN prediction. 
If the initial GNN prediction is not conflict-free, the postprocessing step repairs the remaining conflicts and may introduce new colors when necessary, ensuring that the final output is a \emph{proper coloring}.
The model parameters are optimized using AdamW with an initial learning rate of $1\times10^{-4}$ and a cosine annealing schedule. 
Each model is trained for 2000 epochs with a batch size of 32. 
The Potts loss weight is set to $J=1$, the color-count regularization coefficient to $\beta=0.1$, and the softmax temperature to $\tau=0.3$.

For performance evaluation, we compare the proposed GNN-based method with the DSatur algorithm and the Potts-based unsupervised baseline on the same test instances. 
For each user number $K$ and memory ratio $M/N$, 50 independent conflict graph instances are generated for testing. 
All reported color numbers correspond to the final \emph{proper coloring} after postprocessing; see Section~\ref{sec-postprocessing} for details. 
For the proposed GNN and the Potts-based baseline, the reported runtimes include both the forward/optimization stage and the postprocessing stage, while the DSatur algorithm directly outputs a proper coloring.

All experiments are conducted on a workstation equipped with an NVIDIA GeForce RTX 4090 GPU, and all reported runtimes are measured in the same software and hardware environment to ensure a fair comparison across different methods.

\begin{table}
\centering
\caption{Hyperparameter settings}
\label{tab-training-config}
\begin{tabular}{lc}
\toprule
Parameter & Value \\
\midrule
Number of GNN layers & 3 \\
Hidden dimension & 128 \\
Optimizer & AdamW \\
Initial learning rate & $1\times10^{-4}$ \\
Learning rate schedule & Cosine annealing \\
Batch size & 32 \\
Training epochs & 2000 \\
Potts loss weight $J$ & 1 \\
Color-count regularization $\beta$ & 0.1 \\
Softmax temperature $\tau$ & 0.3 \\
\bottomrule
\end{tabular}
\end{table}

\subsection{Evaluation of the proposed framework}

In this section, we evaluate both the complexity and transmission load of the proposed framework under different numbers of users. All evaluations are conducted under the across user number training setting described in Section~\ref{sec-dataset}. 
The MACC delivery problem under arbitrary user-cache access topology has not been widely studied from a learning-based perspective; consequently, no existing learning-based methods are specifically designed for this problem. 
Therefore, we assess the proposed framework by comparing it with representative traditional graph coloring methods as well as a recent learning-based approach. 
Several learning-based graph coloring methods have been proposed, as described in Section~\ref{sec-graph-coloring}. These methods primarily differ in their choice of backbone architectures, such as GDN, MCGNN or GIN, while generally following a similar learning paradigm. 
To focus on evaluating the proposed framework for the delivery phase of MACC systems, we adopt the Potts-based method in~\cite{schuetz2022potts} as a representative learning-based baseline. 
This method explicitly produces \emph{proper colorings}, i.e., color assignments with no conflict vertices, which ensures the correct delivery phase. 
Other learning-based approaches may not guarantee proper colorings or may prioritize minimizing the number of colors without avoiding conflicts, which is incompatible with MACC delivery requirements.
Specifically, our evaluation focuses on the following comparisons.
\begin{itemize}
    \item \textit{DSatur algorithm vs. IC converse bound:} assessing how closely the classical DSatur algorithm approaches the theoretical IC converse bound;
    \item \textit{Proposed framework vs. DSatur algorithm:} comparing the transmission load and computational complexity of the proposed framework relative to the DSatur algorithm;
    \item \textit{Proposed framework vs. Potts-based method~\cite{schuetz2022potts}:} comparing the transmission load and computational complexity of the proposed framework relative to a recent learning-based method that also produces \emph{proper colorings}.
\end{itemize}
Here, the transmission load is evaluated through the number of colors in the delivery-phase graph coloring formulation, since each color corresponds to one coded transmission group. Accordingly, a smaller number of colors implies a lower delivery transmission load, while the runtime reflects the computational complexity of the corresponding method.

\paragraph{Evaluation settings}
The configurations of the MACC system used in the experiments are summarized in Table~\ref{tab-exp-config}. 

\begin{table}
\centering
\caption{Evaluation settings}
\label{tab-exp-config}
\begin{tabular}{lc}
\toprule
Parameter & Setting \\
\midrule
Number of users $K$ & $\{6,8,10,14,17,19\}$ \\
Number of cache-nodes $\Lambda$ & $10$ \\
Memory ratio $M/N$ & $\{0.1,0.2,\ldots,0.9\}$ \\
Training setting & Cross user number $K \in [4,20]$ \\
Generated instances & $2550$ topologies per $(K,\Lambda)$ \\
Evaluation metrics & transmission load, runtime \\
\bottomrule
\end{tabular}
\end{table}

\paragraph{Gap between the DSatur algorithm and the IC converse bound}

\begin{figure}
    \centering
    \includegraphics[width=0.7\textwidth]{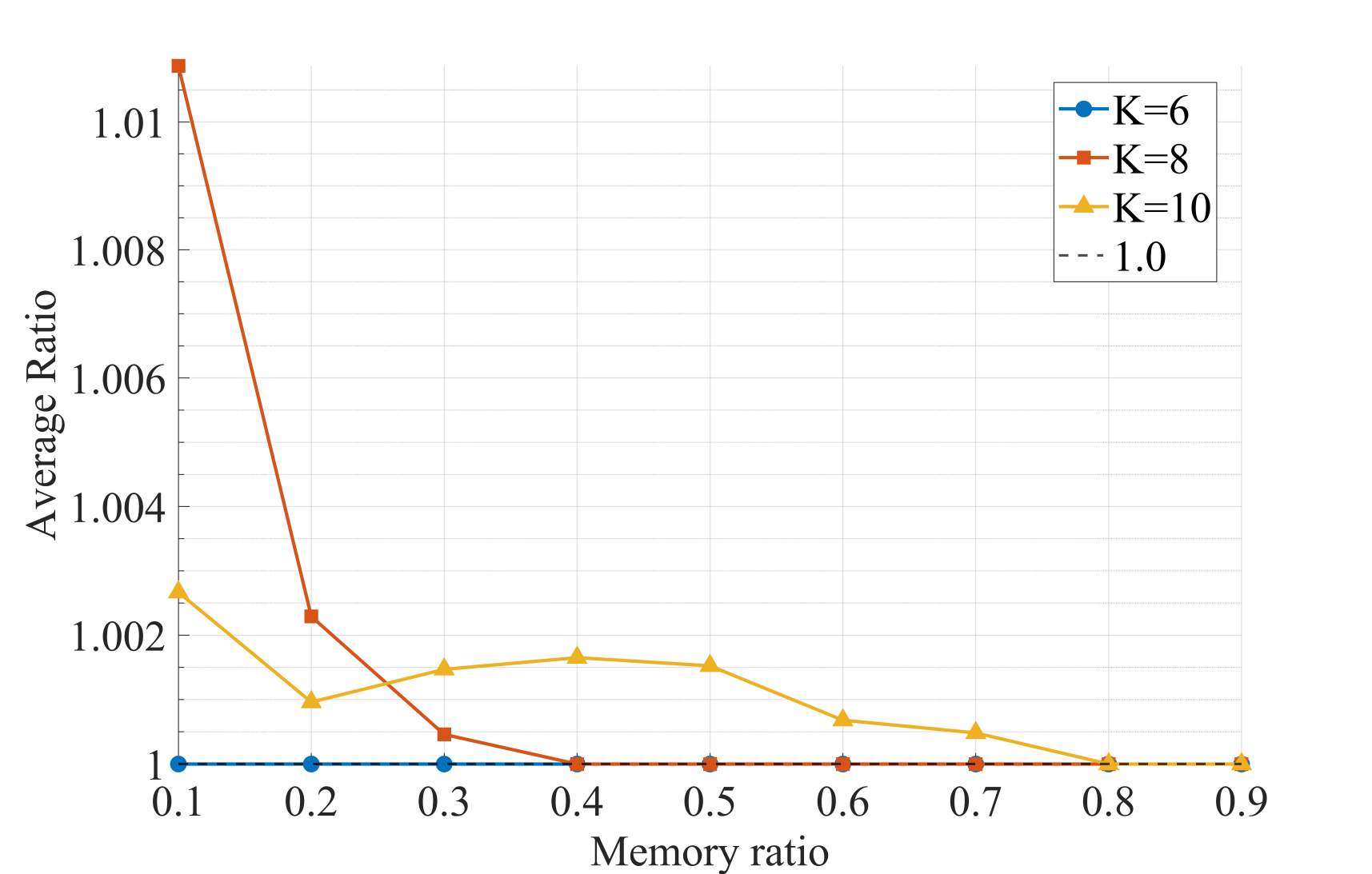}
    \caption{Transmission load ratio between the DSatur algorithm and the IC converse bound.}
    \label{fig-DSvsIC}
\end{figure}

Before evaluating the proposed learning-based framework, we first examine the performance gap between classical graph coloring heuristics and the theoretical lower bound. 
Specifically, we evaluate the ratio between the transmission load achieved by the DSatur algorithm and the IC converse bound.\footnote{In the conflict graph formulation, the number of colors corresponds to the number of transmissions $S$. Under the same cache-node placement, the packetization is fixed across all schemes. Therefore, the color ratio is equivalent to the transmission load ratio.}
Fig.~\ref{fig-DSvsIC} illustrates the ratio $S_{\mathrm{DSatur}}/S_{\mathrm{IC}}$ for different numbers of users. 
For all considered cases $K = 6, 8, 10$, the DSatur algorithm achieves transmission loads extremely close to the IC converse bound. 
For $K=6$, the two bounds are nearly identical with a negligible gap. 
For $K=8$, the maximum relative gap remains below $1.1\%$, while for $K=10$ it further reduces to below $0.3\%$. 
Moreover, at higher memory ratios $M/N$, the DSatur algorithm  almost coincides with the IC converse bound. Overall, the gap does not increase with the number of users. 

\paragraph{Transmission load comparison of the achievable schemes}
We evaluate the transmission load achieved by the proposed GNN-based framework.
The results are compared with the DSatur algorithm and the Potts-based method~\cite{schuetz2022potts}.

\begin{figure}
    \begin{subfigure}{0.45\textwidth}
        \includegraphics[width=\linewidth, height=5cm]{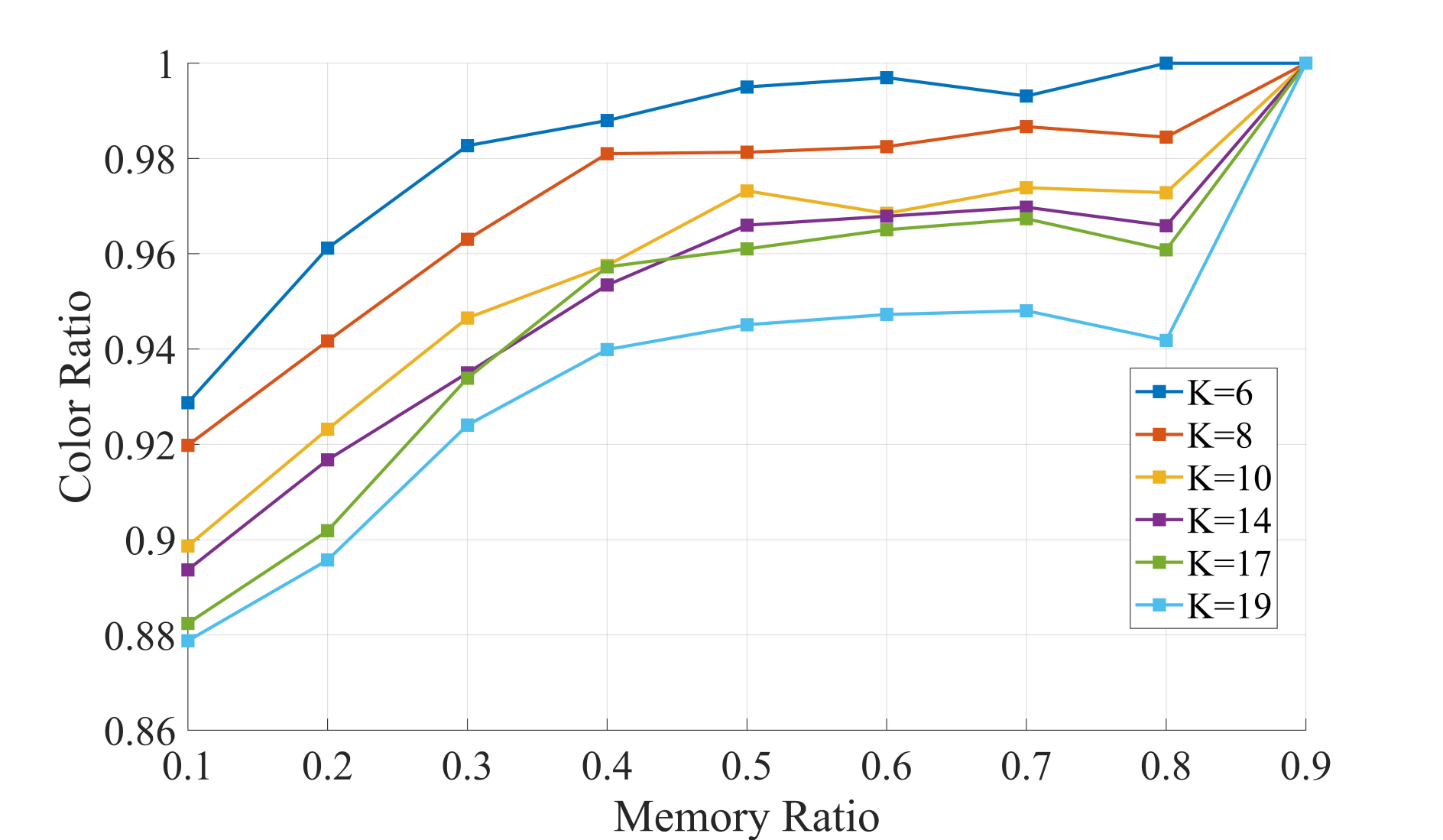}
        \caption{The DSatur algorithm vs. GNN (color ratio).}
        \label{fig-color-DSvsGNN}
    \end{subfigure}
    \hfill
    \begin{subfigure}{0.48\textwidth}
        \includegraphics[width=\linewidth, height=5cm]{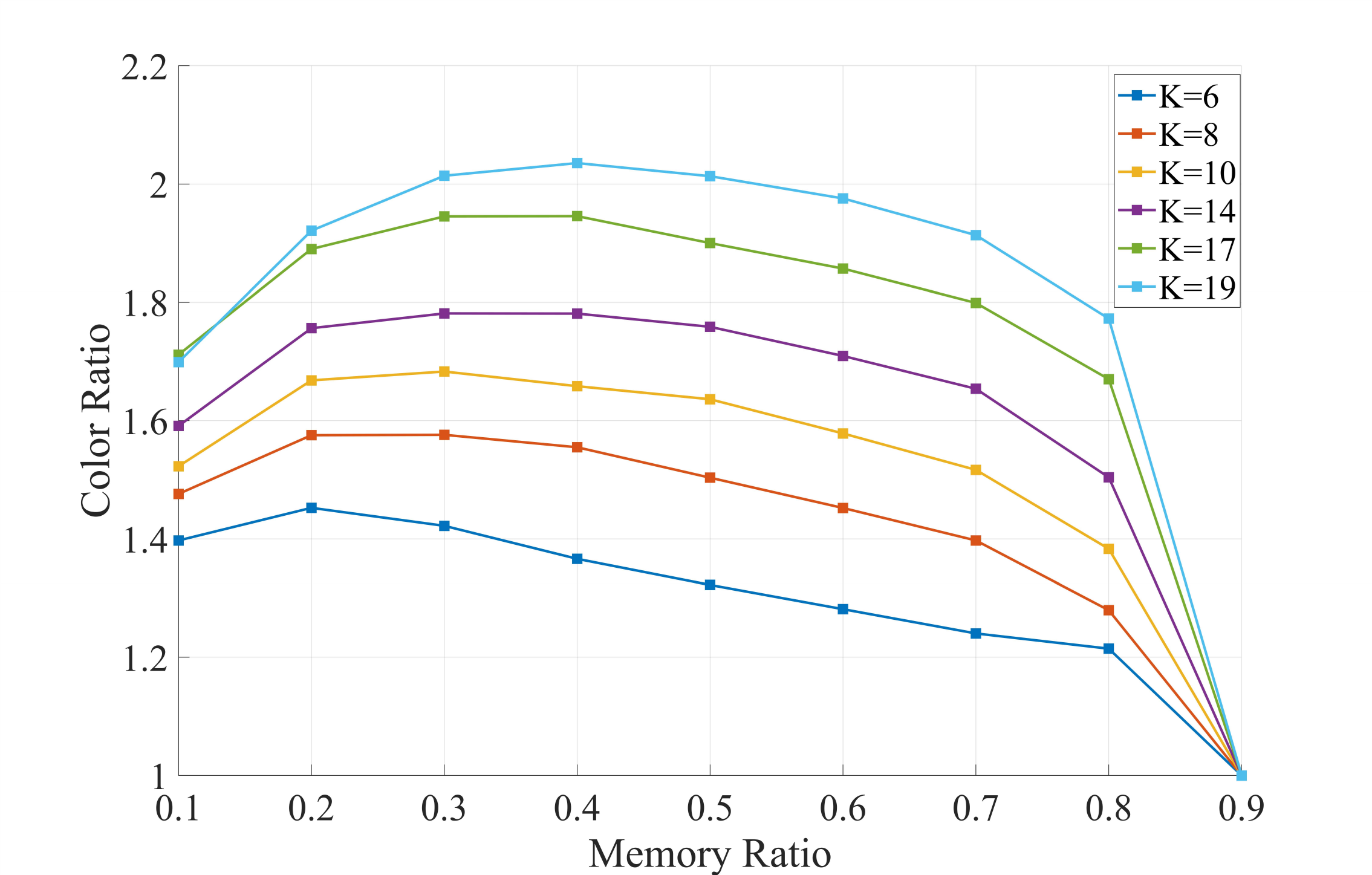}
        \caption{Potts-based vs. GNN (color ratio).}
        \label{fig-color-POTTSvsGNN}
    \end{subfigure}
    \caption{Transmission load comparison of different graph coloring methods under across user number training.}
\end{figure}

\begin{itemize}

\item \textit{Proposed GNN-based framework vs. DSatur algorithm:} 
A single model trained on $K \in [4,20]$ is evaluated on $K \in \{6,8,10,14,17,19\}$. 
Fig.~\ref{fig-color-DSvsGNN} and Table~\ref{tab:performance} show that the ratio $S_{\text{DSatur}}/S_{\text{GNN}}$ remains around $0.88$ to $1$. This indicates that the transmission load achieved by the proposed GNN-based framework remains close to that of the DSatur algorithm. 
Even for larger systems (e.g., $K=19$), the performance gap remains within about $10\%$, demonstrating that the learned policy effectively approximates high-quality graph coloring solutions.


\item \textit{Proposed GNN-based framework vs. Potts-based method~\cite{schuetz2022potts}:} 
Fig.~\ref{fig-color-POTTSvsGNN} shows the ratio $S_{\text{Potts}}/S_{\text{GNN}}$ under across user number training. 
As summarized in Table~\ref{tab:performance}, the ratio remains significantly larger than $1$ across all memory ratios and user numbers, and can exceed $2$ for larger systems such as $K=19$. 
This indicates that the proposed framework consistently requires fewer colors than the Potts-based method while maintaining strong generalization capability.

\end{itemize}

\paragraph{Computational complexity comparison}
We next evaluate the computational complexity of the proposed GNN-based framework by comparing its runtime with the DSatur algorithm and the Potts-based method~\cite{schuetz2022potts}. 

\begin{itemize}

\item \textit{Computational complexity of the DSatur algorithm:} 
The DSatur algorithm has a worst-case time complexity of $\mathcal{O}(|\mathcal{V}|^2 + |\mathcal{E}|)$, which increases rapidly with the number of users $K$ and the density of the induced conflict graph, since the numbers of nodes and edges in the conflict graph grow exponentially with   $K$. 
This motivates the development of efficient learning-based alternatives for large-scale MACC systems.

\item \textit{Proposed GNN-based framework vs. DSatur algorithm:} 
A single GNN model trained on $K \in [4,20]$ is evaluated on $K \in \{6,8,10,14,17,19\}$. 
Fig.~\ref{fig-time-DSvsGNN} and Table~\ref{tab:performance} show that the runtime ratio $T_{\text{DSatur}}/T_{\text{GNN}}$ ranges approximately from $2$ to $30$, indicating that the proposed framework achieves substantial speedup compared with the DSatur algorithm. 
The largest speedup occurs for moderate to large systems (e.g., $K=17$ and $K=19$), highlighting the efficiency advantage of the learned policy for dense conflict graphs.
This indicates that the computational complexity of the GNN is stable, even when a single model is applied across multiple user numbers.

\item \textit{Proposed GNN-based framework vs. Potts-based method~\cite{schuetz2022potts}:} 
Fig.~\ref{fig-time-POTTSvsGNN} and Table~\ref{tab:performance} report the runtime ratio $T_{\text{Potts}}/T_{\text{GNN}}$ across memory ratios and user numbers. 
The ratio ranges from values close to $1$ up to nearly $300$ for larger systems (e.g., $K=19$), and generally increases with the number of users. 
The speedup is less pronounced for very low and high memory ratios, where the conflict graphs are relatively small. 
These results confirm that the proposed framework consistently achieves much lower runtime than the Potts-based method while maintaining robust generalization across different user numbers. 

\begin{figure}
    \begin{subfigure}{0.45\textwidth}
        \includegraphics[width=\linewidth, height=5cm]{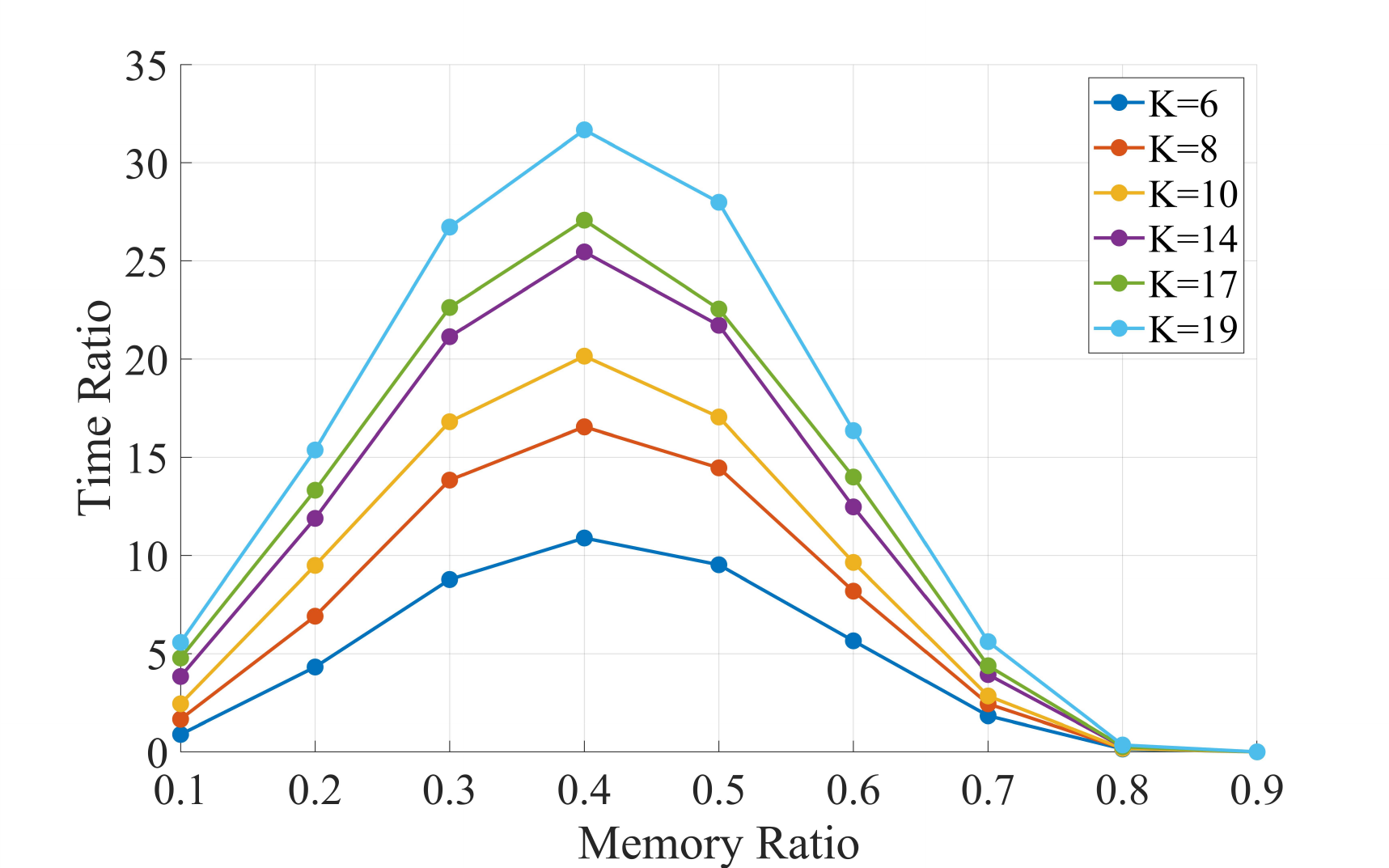}
        \caption{The DSatur algorithm vs. GNN (runtime ratio).}
        \label{fig-time-DSvsGNN}
    \end{subfigure}
    \hfill
    \begin{subfigure}{0.5\textwidth}
        \includegraphics[width=\linewidth, height=5cm]{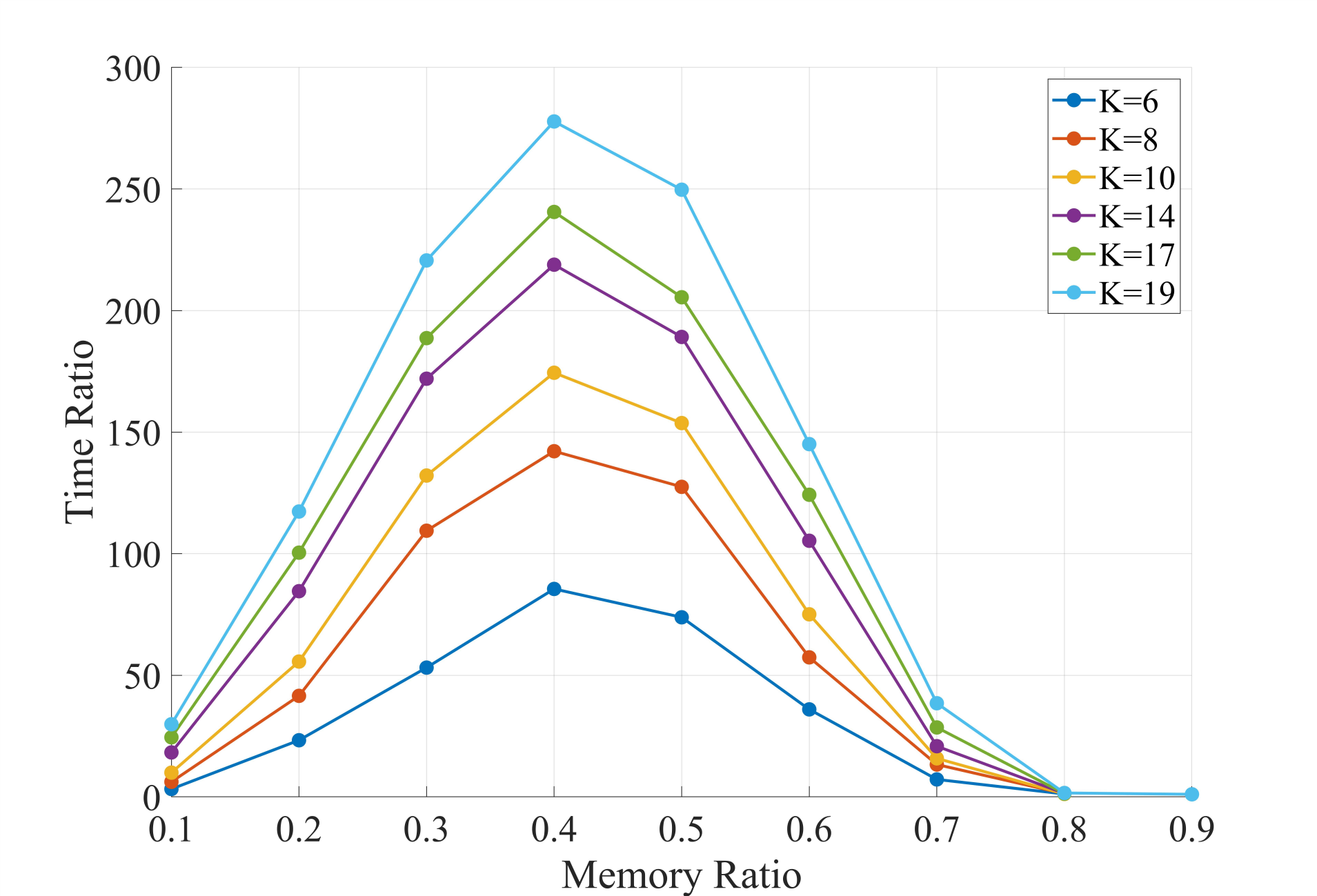}
        \caption{Potts-based vs. GNN (runtime ratio).}
        \label{fig-time-POTTSvsGNN}
    \end{subfigure}
    \caption{Computational complexity comparison of different graph coloring methods under across user number training.}
\end{figure}
\end{itemize}

\begin{table}[h!]
\centering
\caption{Representative transmission load and runtime ratios relative to GNN ($S_\text{method}/S_\text{GNN}$ and $T_\text{method}/T_\text{GNN}$) for different user numbers $K$ across memory ratios $M/N$.}
\label{tab:performance}
\begin{tabular}{c|c|cc|cc}
\toprule
$K$ & $M/N$ & $S_{\text{DSatur}}/S_{\text{GNN}}$ & $S_{\text{Potts}}/S_{\text{GNN}}$ & $T_{\text{DSatur}}/T_{\text{GNN}}$ & $T_{\text{Potts}}/T_{\text{GNN}}$ \\
\midrule
\multirow{9}{*}{10}
& 0.1 & 0.899 & 1.523 & 2.444 & 9.954 \\
& 0.2 & 0.923 & 1.668 & 9.490 & 55.650 \\
& 0.3 & 0.946 & 1.683 & 16.805 & 132.149 \\
& 0.4 & 0.958 & 1.658 & 20.150 & 174.426 \\
& 0.5 & 0.973 & 1.636 & 17.047 & 153.727 \\
& 0.6 & 0.968 & 1.578 & 9.642 & 75.077 \\
& 0.7 & 0.974 & 1.517 & 2.843 & 15.825 \\
& 0.8 & 0.973 & 1.383 & 0.175 & 1.177 \\
& 0.9 & 1.000 & 1.000 & 0.000 & 0.000 \\
\midrule
\multirow{9}{*}{17}
& 0.1 & 0.882 & 1.712 & 4.778 & 24.478 \\
& 0.2 & 0.902 & 1.890 & 13.323 & 100.438 \\
& 0.3 & 0.934 & 1.945 & 22.630 & 188.671 \\
& 0.4 & 0.957 & 1.946 & 27.073 & 240.562 \\
& 0.5 & 0.961 & 1.900 & 22.553 & 205.476 \\
& 0.6 & 0.965 & 1.857 & 13.996 & 124.242 \\
& 0.7 & 0.967 & 1.799 & 4.383 & 28.544 \\
& 0.8 & 0.961 & 1.670 & 0.270 & 1.335 \\
& 0.9 & 1.000 & 1.000 & 0.000 & 0.000 \\
\midrule
\multirow{9}{*}{19}
& 0.1 & 0.879 & 1.699 & 5.575 & 29.861 \\
& 0.2 & 0.896 & 1.921 & 15.375 & 117.353 \\
& 0.3 & 0.924 & 2.014 & 26.720 & 220.618 \\
& 0.4 & 0.940 & 2.035 & 31.668 & 277.775 \\
& 0.5 & 0.945 & 2.013 & 27.986 & 249.699 \\
& 0.6 & 0.947 & 1.976 & 16.354 & 145.074 \\
& 0.7 & 0.948 & 1.914 & 5.613 & 38.503 \\
& 0.8 & 0.942 & 1.772 & 0.352 & 1.586 \\
& 0.9 & 1.000 & 1.000 & 0.004 & 1.053 \\
\bottomrule
\end{tabular}
\end{table}

\paragraph{Ablation study}  
To assess the impact of the neighbor-enhanced message-passing module in the proposed framework (Section~\ref{sec-GNN-train}), we compare the full model with a baseline that replaces this module by a simple linear input projection. Table~\ref{tab:ablation} reports the transmission load ratio $S_{\text{DSatur}}/S_{\text{GNN}}$, where a larger ratio indicates performance closer to the DSatur algorithm. For $K=17$, the full model consistently outperforms the baseline across the considered memory ratios. In all cases, introducing the neighbor-enhanced message-passing module improves the transmission load ratio $S_{\text{DSatur}}/S_{\text{GNN}}$, indicating that exploiting neighborhood information leads to better color assignments and lower transmission load. Similar trends are observed for other system settings.

\begin{table}
\centering
\caption{Ablation study of the proposed GNN framework with $K=17$ and $M/N\in\{0.1,0.2,0.3\}$.}
\label{tab:ablation}
\begin{tabular}{lcc}
\toprule
Variant & $M/N$ &  $S_{\text{DSatur}}/S_{\text{GNN}}$ \\
\midrule
Baseline GNN & 0.1 & 0.869 \\
\textbf{Full model (+Neighbor-enhanced message passing)} & 0.1 & \textbf{0.882} \\
\midrule
Baseline GNN & 0.2 & 0.894 \\
\textbf{Full model (+Neighbor-enhanced message passing)} & 0.2 & \textbf{0.902} \\
\midrule
Baseline GNN & 0.3 & 0.927 \\
\textbf{Full model (+Neighbor-enhanced message passing)} & 0.3 & \textbf{0.934} \\
\bottomrule
\end{tabular}
\end{table}

\subsection{Comparison between greedy converse bound and IC converse bound} \label{sec-compare-greIC}
\begin{figure}
    \centering
    \begin{subfigure}[t]{0.48\textwidth}
        \includegraphics[width=\linewidth, height=6cm]{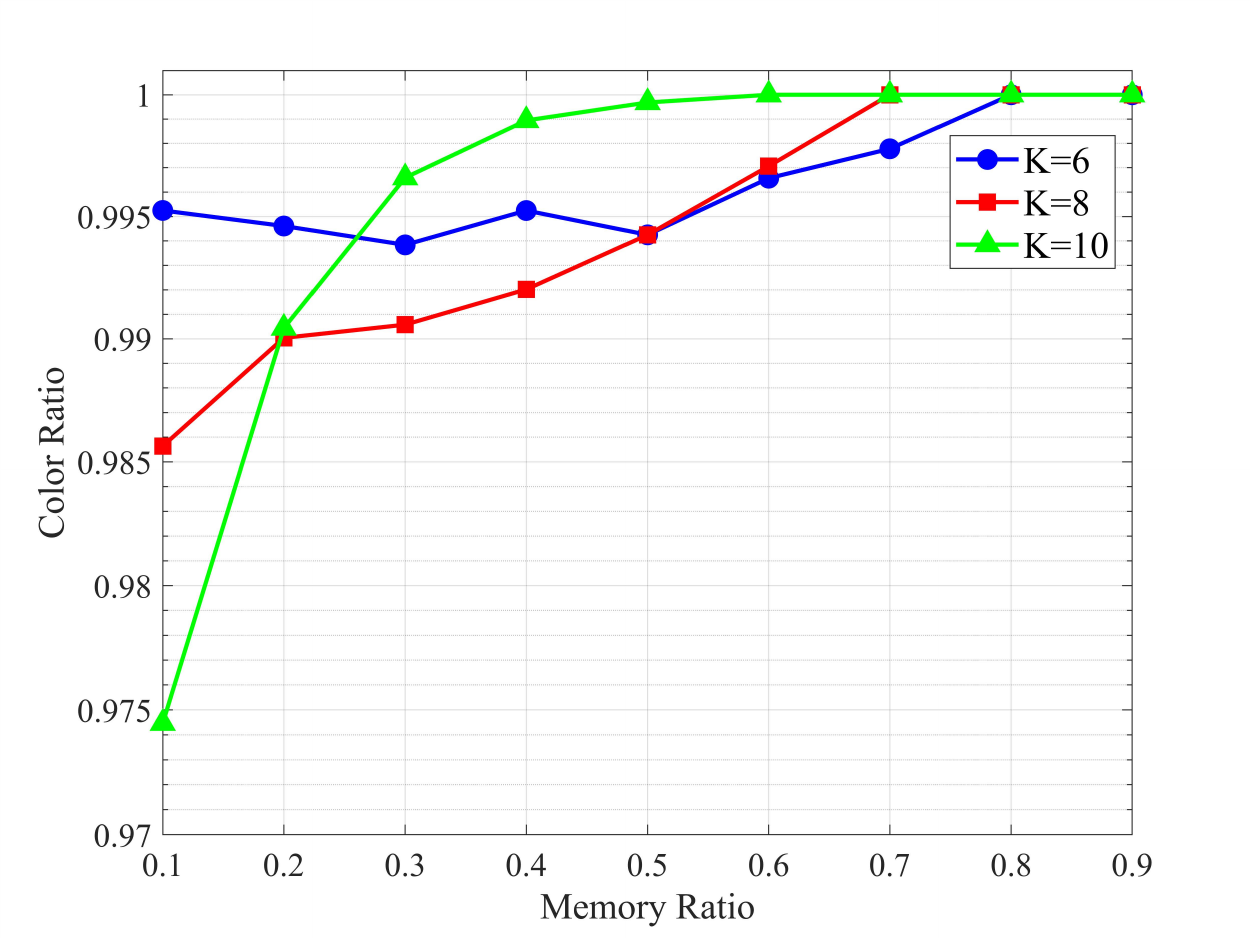}
        \caption{Greedy converse bound ratio relative to the IC converse bound.}
        \label{fig-greedy-color}
    \end{subfigure}
    \hfill
    \begin{subfigure}[t]{0.48\textwidth}
        \includegraphics[width=\linewidth, height=6cm]{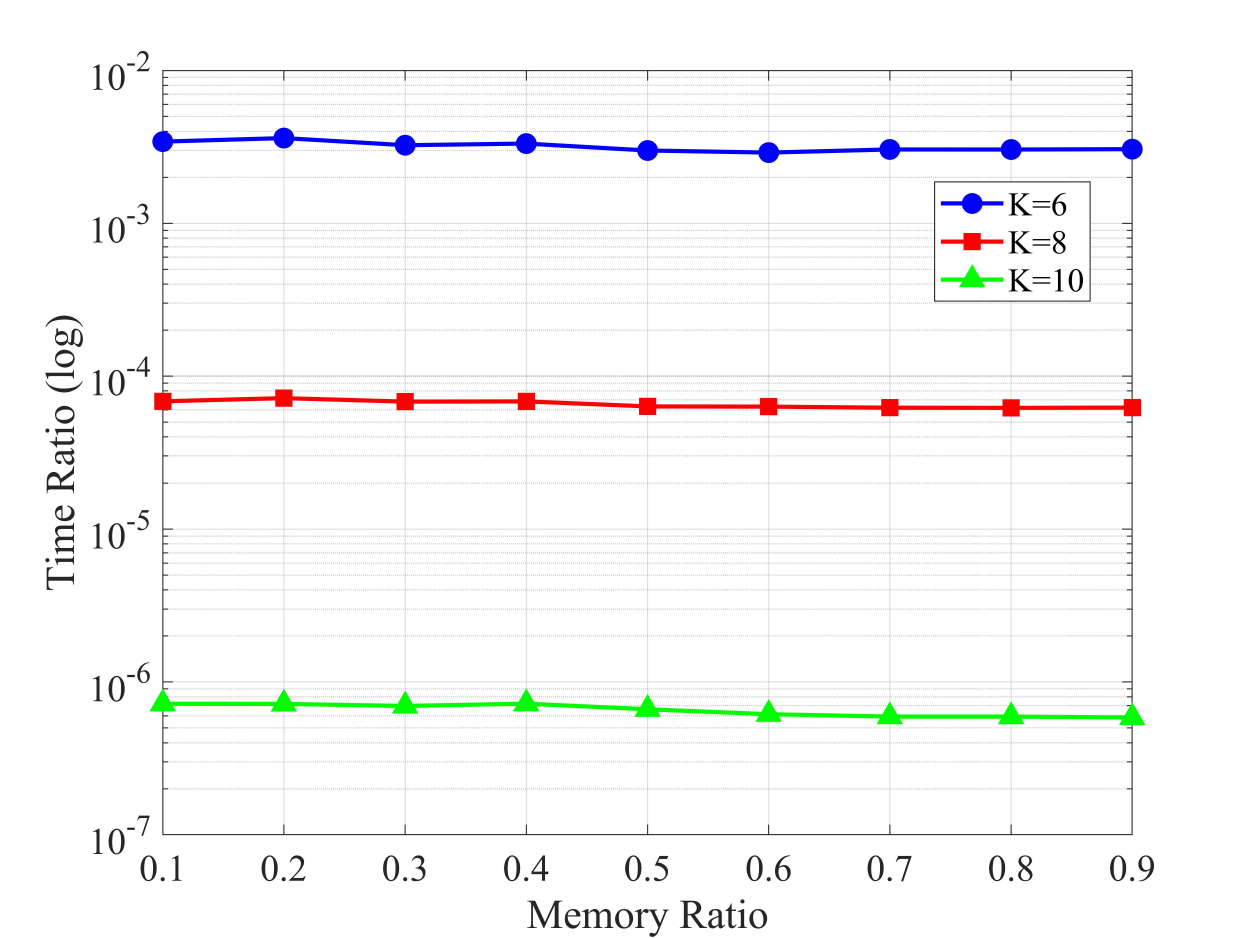}
        \caption{Greedy converse bound runtime ratio relative to the IC converse bound.}
        \label{fig-greedy-time}
    \end{subfigure}
    \caption{Performance comparison between the proposed greedy converse bound and the IC converse bound for different numbers of users.}
    \label{fig:converse comparison}
\end{figure}

As discussed in Section~\ref{sec-converse-IC} and Section~\ref{sec-greedy-converse}, the IC converse bound has a computational complexity that grows factorially with the number of users, i.e., $\mathcal{O}(K!)$, whereas the proposed greedy converse bound has a much lower complexity of $\mathcal{O}(K^2)$. 
In the following, we evaluate the proposed greedy converse bound as a low-complexity alternative to the IC converse bound. Fig.~\ref{fig:converse comparison} compares both the transmission load and runtime across different numbers of users. The results show that the greedy converse bound closely approximates the IC converse bound, achieving over $97\%$ of the IC converse bound across all 50 tested topologies for each user number.

Meanwhile, the computational runtime is drastically reduced. To better visualize the reduction across different system scales, Fig.~\ref{fig:converse comparison} reports the log-scale runtime ratio $\log_{10}(T_{\text{greedy}}/T_{\text{IC}})$. For $K=6$, $8$, and $10$, the log-ratio is approximately $-2$, $-4$, and $-6$, respectively, indicating several orders of magnitude reduction in runtime compared with the IC converse bound.
Table~\ref{tab-greedy-ic} provides a detailed breakdown for a representative system with $K=10$ users and $\Lambda=10$ cache-nodes. The results show that the greedy converse bound achieves more than $97\%$ of the IC converse bound across all memory ratios, and exactly matches the IC converse bound for $M/N \ge 0.6$.


\begin{table}
\centering
\caption{Comparison between the greedy converse bound and the IC converse bound for $K=10$ users and $\Lambda=10$ cache-nodes.}
\label{tab-greedy-ic}
\begin{tabular}{c c c}
\hline
Cache memory ratio
& Load ratio $\left(\frac{S_{\text{greedy}}}{S_{\text{IC}}}\times 100\%\right)$
& Runtime ratio $\left(\log_{10}\frac{T_{\text{greedy}}}{T_{\text{IC}}}\right)$ \\
\hline
$0.1$ & $97.446\%$ & $-6.14$ \\
$0.2$ & $99.044\%$ & $-6.14$ \\
$0.3$ & $99.659\%$ & $-6.16$ \\
$0.4$ & $99.894\%$ & $-6.14$ \\
$0.5$ & $99.968\%$ & $-6.18$ \\
$0.6$ & $100.000\%$ & $-6.21$ \\
$0.7$ & $100.000\%$ & $-6.23$ \\
$0.8$ & $100.000\%$ & $-6.23$ \\
$0.9$ & $100.000\%$ & $-6.23$ \\
\hline
\end{tabular}
\end{table}

\subsection{Discussion}

The experimental results provide several quantitative insights into the performance and scalability of the proposed framework.

\begin{itemize}

\item \textit{The DSatur algorithm as a benchmark:} 
The DSatur algorithm achieves a transmission load ratio $S_{\mathrm{DSatur}}/S_{\mathrm{IC}}$ within approximately $1.1\%$ of the IC converse bound, and for higher memory ratios ($M/N \geq 0.4$) it almost coincides with the converse bound. In addition, the gap to the IC converse bound does not increase noticeably as the number of users grows. This confirms that the DSatur algorithm serves as a strong non-learning benchmark for MACC graph coloring. However, its worst-case computational complexity is $\mathcal{O}(|\mathcal{V}|^2 + |\mathcal{E}|)$, and its runtime increases rapidly as the induced conflict graph becomes larger and denser, which motivates the development of efficient learning-based alternatives for larger MACC systems.

\item \textit{Performance of the proposed GNN-based framework:} 
Compared with the DSatur algorithm, the proposed GNN-based framework achieves transmission load ratios $S_{\text{DSatur}}/S_{\text{GNN}}$ ranging from approximately $0.88$ to $0.97$ for lower memory ratios ($M/N \in [0.1,0.5]$) and from about $0.95$ to $1.0$ for higher memory ratios ($M/N \in [0.6,0.9]$) across all considered user numbers (see Table~\ref{tab:performance}). Even for larger systems such as $K=19$, the ratio remains above about $0.88$, indicating that the performance gap relative to the DSatur algorithm is within about $12\%$. These results demonstrate that the proposed framework can effectively approximate the DSatur algorithm while generalizing well across different user-cache access topologies and user numbers in MACC systems.

\item \textit{Substantial computational complexity reduction:} 
Compared with the DSatur algorithm, the proposed GNN-based framework reduces runtime by a factor ranging from about $2$ to more than $30$ for low and moderate memory ratios ($M/N \in [0.1,0.7]$), with larger reductions generally observed for larger systems (Table~\ref{tab:performance}). The improvement is particularly pronounced for $M/N \in [0.2,0.6]$, where the runtime reduction is typically above one order of magnitude and can reach $30$ for $K=19$. This is because the induced conflict graphs are more challenging in this regime, making the DSatur algorithm more computationally expensive. For higher memory ratios, especially $M/N=0.8$ and $0.9$, the conflict graphs become easier to color, and the runtime advantage of the GNN becomes much less pronounced and may even disappear in some cases. Overall, these results confirm that the proposed framework significantly reduces computational complexity while maintaining transmission load performance close to that of the DSatur algorithm.

\item \textit{Advantage over existing learning-based approaches:} 
Compared with the Potts-based method in~\cite{schuetz2022potts}, the proposed framework achieves both lower transmission load and substantially lower runtime. The transmission load ratio $S_{\text{Potts}}/S_{\text{GNN}}$ from about 1.4 to slightly above 2 for $M/N \in [0.1,0.8]$, and generally increases with the number of users, indicating that the Potts-based method requires significantly more transmissions. In terms of computational complexity, the runtime ratio $T_{\text{Potts}}/T_{\text{GNN}}$ ranges from about $10$ to $300$, with the largest gains again observed for low and moderate memory ratios ($M/N \in [0.2,0.6]$). These results show that the proposed framework achieves substantially lower runtime than the Potts-based method while maintaining strong transmission load performance.

\item \textit{Performance of the greedy converse bound:} 
The IC converse bound provides a theoretical lower bound on the transmission load, but its computational complexity grows factorially with the number of users, i.e., $\mathcal{O}(K!)$, which makes it infeasible for larger systems. The proposed greedy converse bound closely approximates the IC converse bound, achieving over $97\%$ of the converse-bound performance across all tested topologies for various user numbers, and coinciding exactly with the IC converse bound at high memory ratios. Meanwhile, the greedy converse bound drastically reduces computational time with complexity $\mathcal{O}(K^2)$. The log-scale runtime ratio $\log_{10}(T_{\text{greedy}}/T_{\text{IC}})$ reaches approximately $-2$, $-4$, and $-6$ for $K=6$, $8$, and $10$, respectively, demonstrating orders-of-magnitude reductions in runtime while maintaining high accuracy. These results confirm that the greedy converse bound provides an efficient and reliable alternative to the IC converse bound, particularly for large-scale MACC systems.

\end{itemize}

Overall, these results demonstrate that the proposed GNN-based graph coloring framework provides an efficient and scalable solution for large-scale MACC systems with arbitrary user-cache access topology.

\section{Conclusion}
This paper studied MACC systems with arbitrary user-cache access topology. Focusing on the delivery phase under the MN cache placement, we first proposed a universal graph-based formulation that transforms the delivery problem into a graph coloring problem. 
Within this framework, the classical DSatur algorithm was shown to achieve transmission loads close to the IC converse bound, serving as a strong and practical performance benchmark. 
To address the computational challenges associated with large-scale graphs, we further developed a GNN-based graph coloring framework that learns efficient coloring policies directly from graph structures. 
The proposed learning-based approach achieves a transmission load comparable to the DSatur algorithm,  while substantially reducing computational time and exhibiting strong generalization across different access topologies and  numbers of users. 
In addition, we proposed a low-complexity greedy approximation of the IC converse bound for the MACC systems  with arbitrary access topology. The greedy converse bound closely matches the IC converse bound while offering orders-of-magnitude reductions in computational complexity. 
Future work includes the joint optimization of cache placement and delivery strategies, as well as extensions to dynamic or time-varying user-cache access topologies.

\bibliographystyle{IEEEtran}
\bibliography{reference}
\end{document}